\documentclass[aps,prl,twocolumn,superscriptaddress]{revtex4-2}
\usepackage{amsmath,amssymb}
\usepackage{graphicx}
\usepackage{subfigure}
\usepackage{verbatim}
\usepackage{placeins}
\usepackage{amsfonts}
\usepackage{dcolumn}
\usepackage{bm}
\usepackage{color}
\usepackage[colorlinks,citecolor=blue]{hyperref}
\usepackage{dcolumn}
\usepackage{bm}
\usepackage{mathrsfs}
\usepackage{cancel}
\usepackage{multirow}
\usepackage{CJK}
\usepackage{verbatim}
\usepackage{orcidlink}
\usepackage[normalem]{ulem}

\newcommand{\macro}[1]{\texttt{\textbackslash#1}}
\newcommand{\m}[1]{\macro{#1}}

\usepackage{xcolor}
\definecolor{orange}{rgb}{1,0.5,0}
\definecolor{darkblue}{rgb}{0.,0.,0.6}
\definecolor{darkred}{rgb}{0.5,0.,0.}

\begin{document}

\title{
Fractionalized XY$^*$ criticality in bilayers with opposite magnetic flux
}

\author{Liangdong Hu}
\altaffiliation{The two authors contributed equally to this work.}
\affiliation{School of Physics and Electronics, Hunan University, Changsha 410082, China}

\author{Matthew Yusuf} 
\altaffiliation{The two authors contributed equally to this work.}
\affiliation{School of Physics and Astronomy, University of Leeds, Leeds LS2 9JT, United Kingdom}

\author{Jiangyuan Qian}
\affiliation{Department of Particle Physics and Astrophysics, Weizmann Institute of Science, Rehovot 7610001, Israel
 }

\author{Chao Han}
\affiliation{Institute of Natural Sciences, Westlake Institute for Advanced Study, Hangzhou 310024, China}

\author{Zlatko Papi\'c}
\email{z.papic@leeds.ac.uk}
\affiliation{School of Physics and Astronomy, University of Leeds, Leeds LS2 9JT, United Kingdom}

\author{Liujun Zou}
\email{lzou@nus.edu.sg}
\affiliation{Department of Physics, National University of Singapore, 117551, Singapore}

\author{W. Zhu}
\email{zhuwei@westlake.edu.cn}
\affiliation{Institute of Natural Sciences, Westlake Institute for Advanced Study, Hangzhou 310024, China}
\affiliation{Department of Physics, School of Science, Westlake University, Hangzhou 310030, China }

\date{\today}

\begin{abstract}
A continuous transition between topological order and 
symmetry breaking provides a natural setting for fractionalized quantum criticality beyond the Landau paradigm.
Here we demonstrate such a transition in a bosonic bilayer at filling $\nu=1/2$ per layer, subject to magnetic fields of equal magnitude and opposite signs. At weak interlayer attraction, the system realizes a fractional quantum spin Hall state formed by two opposite-chirality Laughlin liquids, while stronger attraction produces an interlayer-paired superfluid.  
Using fuzzy-sphere spectroscopy of the transition, we show that the low-energy spectrum organizes into conformal families consistent with XY$^*$ Wilson-Fisher criticality, with the physical superfluid order parameter corresponding to the charge-two primary field. 
We further find evidence for a $\mathbb{Z}_3$ counterpart of the same transition in electronic bilayers at fillings $(1/3,2/3)$, 
relevant to semiconductor quantum Hall experiments.
Our results provide a microscopic realization of anyon condensation in which the destruction of topological order and the onset of conventional long-range order occur together.
\end{abstract}

\maketitle

{\bf \em Introduction.---} The quantum spin Hall effect, made up of two time-reversal-related counterparts with opposite chiralities, exhibits a vanishing electrical Hall response but a quantized (pseudo-)spin Hall response~\cite{KaneMele2005,BernevigZhang2006}. Strong interactions enrich this construction into a fractional quantum spin Hall (FQSH) phase~\cite{LevinStern_2009,Maciejko_2015,Levin2012}, which realizes an intrinsic topological order \cite{Wen_book} rather than merely a band insulator.
This FQSH phase has been challenging
to realize in traditional  materials such as double quantum wells because it requires strong but oppositely directed magnetic fields.  Recent advances in moir\'e materials (e.g., MoTe$_2$, WSe$_2$) have opened a door to this elusive phase of matter since two isolated valley Chern bands  naturally experience opposite magnetic flux, where a series of candidate FQSH states have recently been observed~\cite{kang2024,YPWang2026}. 

When the layer degree of freedom acts as a controllable pseudospin, interlayer interaction may favor a coherent bound state of the two layers, giving rise to superfluidity (SF)~\cite{Halperin1983,annurevEisenstein,MacDonald2004}. 
Double-layers therefore can exhibit two sharply different forms of quantum matter: long-range-entangled topological orders like FQSH and spontaneous phase-coherent SF. Their competition raises the question: Can an FQSH phase be converted  into an SF via a continuous quantum phase transition?
A natural route for reconciling topology and spontaneous phase coherence is anyon condensation~\cite{Bais2009,KONG2014,Burnell_2018}: the condensation of a bosonic interlayer bound state of fractional excitations can simultaneously destroy topological order and establish SF coherence~\cite{ChenYang2012,YahuiZhang2023}. 
Nevertheless, verifying this scenario is challenging, since what becomes critical are usually not local fields. Existing attempts have been limited to critical correlators around the transition point~\cite{Isakov_2012,YahuiZhang2023}, which do not fully resolve the fractionalized nature of criticality.

Recently, fuzzy sphere regularization \cite{Zhu2023FuzzySphere,Fuzzysphere_review2026} has been introduced to the study of $(2+1)$D phase transitions and used to uncover potential conformal field theories (CFTs) \cite{Hu_2023,taylor_25,Voinea_2026,Voinea_2025,Zhou_2026,Zhou:2025rmv,huffman_26,Yang_25,Zhou_24,Feng:2026iii,Hu_2024,han:2023,Guo:2025odn,dey_2026,dey_2026_2,stergiou_26,Zhou_2025,Lauchli_2025}. It has been applied to many strongly coupled critical theories, such as Wilson-Fisher CFTs \cite{Zhu2023FuzzySphere,Hu_2023,Lauchli_2025,taylor_25,Yang_25,han:2023,Guo:2025odn,dey_2026,dey_2026_2,cruz_25,miro_2025,fan_2025}, surface/defect CFTs \cite{Feng:2026iii,Hu_2024,dedushenko_24,Zhou_2024}, and deconfined criticality \cite{Zhou_24,Chen:2023xjc,Yang_25_DQCP,huffman_26}. Very recently, this framework was successfully applied to unveil topological phase transitions, such as the Chern-Simons-Higgs transition from a trivial insulator to a chiral spin liquid \cite{Zhou:2025rmv}, and quantum Hall transitions involving non-Abelian states \cite{Voinea_2026,Zhou_2026}. 

In this paper, we employ the  fuzzy-sphere regularization to investigate the phase transition from an FQSH phase to an SF in a bosonic bilayer with opposite magnetic flux, and to probe its conformal nature. We first map out the phase diagram and show that it includes an FQSH phase with doubled-semion topological order and a paired SF phase. We numerically demonstrate the closing of the gap and extract the critical operator content of the CFT via the state-operator correspondence. 
We establish fingerprints of an emergent conformal symmetry, and provide evidence for a fractionalized XY$^*$ transition whose neutral local spectrum is XY-like but whose physical order parameter retains the imprint of fractionalization. In End Matter, we further investigate an electronic quantum Hall bilayer with Coulomb interactions and finite layer width. Its critical spectrum and exciton correlations provide evidence for a $\mathbb{Z}_3$ realization of XY$^*$ criticality, placing our conformal analysis in an experimentally relevant setting.

\begin{figure}[t]
    \centering
    \includegraphics[width=\columnwidth]{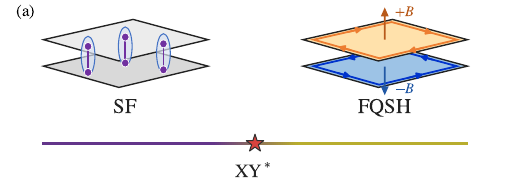}
    \includegraphics[width=\columnwidth]{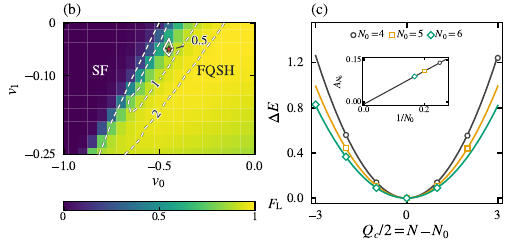}
    \caption{
    (a) Schematic of the transition between the paired SF (left) and FQSH phase (right).
    (b) Laughlin fidelity $F_L$ for $N=7$ (color).
    White contours show the spectral cost $f_{\mathrm{cost}}^{(3)}=0.5,1,2$, defined in
    Eq.~\eqref{eq:spectral_cost}, for the projected $N=8$ data.
    The high-fidelity region identifies the FQSH phase, while the low-fidelity region at stronger attraction corresponds to the paired SF.
    The star marks the cost minimum at $(-0.45,-0.05)$, with $f_{\mathrm{cost}}^{(3)}=0.20515$.
    (c) Lowest $U(1)_c$ excitation energy $\Delta E$ versus $Q_c/2=N-N_0$ in the SF phase at $(v_0,v_1)=(-0.8,-0.05)$, where $N_0$ is the reference particle number per layer.
    Solid curves are quadratic fits, $\Delta E=A_{N_0}(Q_c/2)^2$.
    Inset: fitted curvature $A_{N_0}$ versus $1/N_0$.
   }
    \label{fig:laughlin-fidelity}
\end{figure}

{\bf \em Theoretical perspective.---}We  consider a bosonic bilayer system depicted in Fig.~\ref{fig:laughlin-fidelity}(a).
In the decoupled limit, each layer forms a 
$\nu=1/2$ Laughlin state, but  with opposite chiralities. This is described by a double-semion theory:
\begin{equation}
  \mathcal{L}_{\mathrm{FQSH}}
  =
  \frac{2}{4\pi}a_{\uparrow}d a_{\uparrow}
  -\frac{2}{4\pi}a_{\downarrow}d a_{\downarrow}
  +\frac{1}{2\pi}A_{\uparrow}d a_{\uparrow}
  +\frac{1}{2\pi}A_{\downarrow}d a_{\downarrow},
  \label{eq:topological_action}
\end{equation}
where $a_{\uparrow,\downarrow}$ are dynamical Chern--Simons gauge
fields, $A_{\uparrow,\downarrow}$ probe the separately conserved
layer charges, and
$a d b\equiv\epsilon^{\mu\nu\lambda}a_{\mu}\partial_{\nu}b_{\lambda}$.
The $\uparrow$-layer ($\downarrow$-layer) gives a semion $s$ (anti-semion $\bar s$) with self-statistics $i$ ($-i$) and charge 1/2 under $A_\uparrow$ ($A_\downarrow$). The fusion product $s\bar s$ is another anyon that has bosonic self-statistics and charge 1/2 under both $A_\uparrow$ and $A_\downarrow$.

In the presence of strong interlayer attraction, the system naturally favors 
the bound state $s\bar s$. 
Close to the transition point, the minimal Landau-Ginzburg-Chern-Simons theory takes the form~\cite{ChenYang2012,YahuiZhang2023}
\begin{equation}
  \begin{split}
    \mathcal{L}_{\mathrm{LGCS}}
    &={}
    \left|(\partial_{\mu}-i a_{s,\mu})\phi\right|^2
    +r|\phi|^2+u|\phi|^4
    \\
    &+\frac{1}{2\pi}a_c d a_s
    +\frac{1}{2\pi}A_c d a_c
    +\frac{1}{2\pi}A_s d a_s+\cdots .
  \end{split}
  \label{eq:LGCS_action}
\end{equation}
where 
$a_c=a_{\uparrow}+a_{\downarrow}$,
$a_s=a_{\uparrow}-a_{\downarrow}$ and, analogously,
$A_c=(A_{\uparrow}+A_{\downarrow})/2$ and
$A_s=(A_{\uparrow}-A_{\downarrow})/2$. 
Accordingly, $\phi$ has internal gauge charges
$(q_c^{\mathrm{g}},q_s^{\mathrm{g}})=(0,1)$, despite carrying the
physical charges $(Q_c,Q_s)=(1,0)$. It therefore couples minimally to
$a_s$, but not to $a_c$. Due to the mutual Chern-Simons term $\frac{1}{2\pi}a_cda_s$, a $2\pi$ flux of $a_c$ is attached to a quantum of $\phi$, and this composite is the anyon $s\bar s$.

To check the validity of the effective theory in Eq.~\eqref{eq:LGCS_action}, note that it leads to a topological FQSH phase for $r>0$. Specifically, in this case, $\phi$ is massive and Eq.~\eqref{eq:LGCS_action} reduces to the $U(1)_2\times U(1)_{-2}$ topological field theory  in Eq.~\eqref{eq:topological_action}. The fields $a_c$
and $a_s$ are nonpropagating, and the semion and antisemion remain
deconfined. On the other hand, for $r<0$, $\phi$ develops a nonzero amplitude and induces the condensation of $s\bar s$.
Writing $\phi=v e^{i\theta}$ produces
$\rho_s(d\theta-a_s)^2$, so $a_s$ absorbs $\theta$ and becomes
massive due to the Higgs effect. This Higgs effect also renders the Chern-Simons term $\frac{1}{2\pi}a_cda_s$ unimportant,  eliminating the
double-semion topological order. Integrating out the massive $a_s$ yields a Maxwell term,
\begin{align}
\mathcal{L}_{\mathrm{SF}}[a_c]\propto(d a_c)^2/\rho_s.
\end{align}
In $2+1$ dimensions, its single gapless mode is the dual
representation of the SF Goldstone mode \cite{Peskin1978, Dasgupta1981}. Since $s\bar s$ carries charge under $A_c$ but not $A_s$, the $U(1)_c$ symmetry related to the total boson number conservation is spontaneously broken, while the $U(1)_s$ symmetry related to the conservation of the difference between the boson numbers of the two layers is preserved. Note that $s\bar s$ is an anyon while the bound state of two such anyons is a physical interlayer exciton, which is the physical order parameter of this symmetry-breaking phase and carries charge 2 under $A_c$. Therefore, the effective theory Eq.~\eqref{eq:LGCS_action} describes a paired SF where $U(1)_c$ is spontaneously broken to its $\mathbb{Z}_2$ subgroup when $r<0$.

At the transition point $r=0$, integrating out $a_c$ in Eq.~\eqref{eq:LGCS_action} imposes $a_s=-A_c$ up to a gauge transformation. The critical scalar consequently corresponds as a particle of total physical charge 1, which is consistent with the charge of $s\bar s$, and Eq.~\eqref{eq:LGCS_action} simplifies to
\begin{equation}
  \begin{split}
    \mathcal{L}_{\mathrm{XY}^*}
    ={}&
    \left|(\partial_{\mu}+i A_{c,\mu})\phi\right|^2
    +r|\phi|^2+u|\phi|^4-\frac{1}{2\pi}A_cdA_s.
  \end{split}
  \label{eq:xy_star_action}
\end{equation}
The  $\frac{1}{2\pi}A_c dA_s$ term retains the quantized spin-Hall information of the adjacent FQSH phase but introduces no independent gapless gauge fluctuation. 
Eq.~\eqref{eq:xy_star_action}, known as the XY* theory, is closely related to the 3D $O(2)$ XY Wilson-Fisher critical theory, but with a crucial difference: the entire local operator spectrum of the XY* theory is only the charge-even local operator spectrum of the XY theory, while the charge-odd spectrum of the XY theory is absent in the XY* theory as $\phi$ is coupled to a dynamical gauge field. 

It is worth noting that the present realization of the XY* theory is quite distinct from the transition identified in the  $\mathbb{Z}_2$ quantum spin liquid \cite{Isakov_2012,Schuler_2023}.  The reason is, although the XY* theory can be generally obtained by gauging a $\mathbb{Z}_2$ symmetry of the XY theory, there are multiple possibilities in doing so. First, the $\mathbb{Z}_2$ symmetry to be gauged may or may not be a subgroup of the $U(1)$ symmetry of the original XY theory. If it is a subgroup, then any (gapped or gapless) charge-1 excitation will couple to a $\mathbb{Z}_2$ gauge field and cease to be local, which is exactly what happens in Ref.~\cite{Isakov_2012}. However, in our case, the original bosons in both layers carry charge-1 and they remain as (gapped) local excitations, so the $\mathbb{Z}_2$ symmetry to be gauged is {\it not} a subgroup of the $U(1)$ symmetry of the XY theory. Second, whenever a $\mathbb{Z}_2$ symmetry is gauged, one has to specify whether the system hosts a nontrivial $\mathbb{Z}_2$ symmetry-protected topological order before gauging. If it does (does not), then the gapped side of the XY* theory hosts a double-semion (toric code) topological phase, which is exactly what happens in our case (Ref. \cite{Isakov_2012}) \cite{Levin2012a}, see the Supplemental Material (SM)~\cite{sm} for details.

{\bf \em Model.---}To simulate the FQSH-SF transition, we consider a double-layer fuzzy (Haldane) sphere~\cite{Haldane1983}, subject to monopole magnetic fluxes of $\pm s$ with equal magnitude but opposite signs in the two layers~\cite{Furukawa2014}.
In the lowest Landau level, each layer consists of $N_\phi=2s$  magnetic flux quanta and is filled by $N$ bosons. 
We focus on bosons at half filling, i.e. $\nu=1/2$ and $N_\phi=2N-2$, with opposite chiralities, subject to the following intralayer and interlayer interactions:
\begin{equation}
    \begin{split}
        H =& \int d^2 \boldsymbol{\Omega} \int d^2 \boldsymbol{\Omega}^\prime \, U_{\sigma\sigma}(\boldsymbol{\Omega}-\boldsymbol{\Omega}^\prime)  n_\sigma(\boldsymbol{\Omega})n_{\sigma}(\boldsymbol{\Omega}^\prime)     \\   
    &+\int d^2 \boldsymbol{\Omega} \int d^2 \boldsymbol{\Omega}^{\prime} \, U_{\sigma\bar{\sigma}}(\boldsymbol{\Omega}-\boldsymbol{\Omega}^{\prime})  n_\sigma(\boldsymbol{\Omega})n_{\bar{\sigma}}(\boldsymbol{\Omega}^{\prime}) 
    \end{split}\label{eq:ham}
\end{equation}
where $\boldsymbol{\Omega}=(\theta,\phi)$ is spherical coordinate, $n_\sigma$ is the bosonic density operator in layer $\sigma=\uparrow,\downarrow$. The interactions are parameterized by the Haldane pseudopotentials \cite{Haldane1983}.
The two layers have identical intralayer interactions, with only the
repulsive $V_0^{\mathrm{intra}}=1$ retained. To favor a paired SF, we choose attractive interlayer interactions
with $V_0^{\mathrm{inter}}=v_0<0$ and $V_1^{\mathrm{inter}}=v_1<0$,
setting all higher pseudopotentials to zero.

The Hamiltonian in Eq.~\eqref{eq:ham} is solved 
by projecting onto the lowest Landau level, and expanding the bosonic
fields in monopole harmonics~\cite{WU1976}.
The projected Hamiltonian is provided in the SM~\cite{sm}.
In the exact diagonalization calculation, we fully utilize the symmetries of this model using the FuzzifiED package \cite{ZhouHe2026SO3lver}:
$U(1)_\uparrow\times U(1)_\downarrow$ symmetry due to the conservation of  particle number in each layer, spatial rotational symmetry $SO(3)_r$ on the sphere, and time reversal symmetry $\mathcal{M}_T$. Implementing these symmetries not only speeds up the calculations but also 
helps to distinguish different low-lying excitations by their corresponding symmetry quantum numbers. 

\begin{figure}[tbp]
    \centering
    \includegraphics[width=\columnwidth]{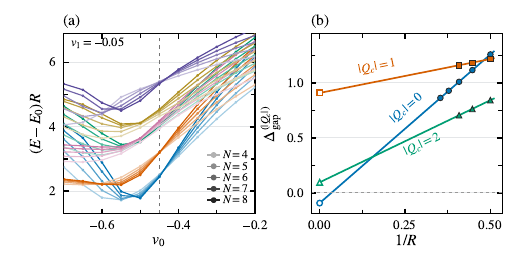}
    \caption{
    (a) Rescaled gaps $(E_n-E_0)R$ after magnetoroton projection
    along $v_1=-0.05$, for $N=4$--$8$, with $R=\sqrt N$.
    Colors distinguish the first two excitations at each $\ell=0,1,2$.
    Shades darken with increasing $N$.
    (b) Neutral and particle-hole-averaged $|Q_c|=1,2$ gaps at $(v_0,v_1)=(-0.45,-0.05)$.
    Open symbols mark extrapolations via the linear fit.
}
    \label{fig:scaled-gap-crossing}
\end{figure}

{\bf \em Numerical results.---}We begin by mapping out the phase diagram defined by the interlayer interaction
parameters $(v_0,v_1)$. At the decoupled point $v_0=v_1=0$, each layer
realizes a bosonic $\nu=1/2$ Laughlin state $|\Psi_{\mathrm L_{\sigma}}\rangle$, and the model ground
state of the FQSH phase is their opposite-chirality product, $|\Psi_{\mathrm {FQSH}}\rangle =|\Psi_{\mathrm L_{\uparrow}}\rangle \otimes |\Psi_{\mathrm L_{\downarrow}}\rangle =|\Psi_{\mathrm L}\rangle \otimes |\overline{\Psi_{\mathrm L}}\rangle  $. 
To track the stability of this phase away
from the decoupled limit, we calculate the fidelity
$  F_L(v_0,v_1)
  =  \left|
    \langle\Psi_{\mathrm{FQSH}}|\Psi_0(v_0,v_1)\rangle
  \right|^2$,
where $|\Psi_0(v_0,v_1)\rangle$ is the  ground state at parameters $(v_0,v_1)$.
Figure~\ref{fig:laughlin-fidelity}(b) shows the $N=7$ fidelity $F_L$ across
the $(v_0,v_1)$ plane. A broad high-fidelity region is
continuously connected to the decoupled point, demonstrating that
the FQSH phase remains stable against moderate interlayer
interactions. In this regime, the ground state retains the
$U(1)_2\times U(1)_{-2}$ topological order of the two
opposite-chirality Laughlin layers. Upon making $v_0$ increasingly
attractive, fidelity decreases rapidly, signaling the loss of
the Laughlin character. The small fidelity and sharp suppression provide a finite-size indicator of the boundary of the FQSH regime.

Strong attraction is expected to yield a SF phase with paired bosons between the two layers. The formation of the SF phase should spontaneously break $U(1)_c$ symmetry, which can be detected from the energy gap,
$\Delta E=E_0(N_0+\Delta N)-E_0(N_0)-\mu_{\mathrm p}\Delta N$,
where $E_0(N)$ is the lowest energy in the balanced sector with $N$ bosons per layer at fixed monopole strength. Here $N_0$ is the reference particle number per layer, $\mu_{\mathrm p}$ is the pair chemical potential, see SM~\cite{sm}. Figure~\ref{fig:laughlin-fidelity}(c) demonstrates that even-$Q_c$ energy gaps are well described by the quadratic form $\Delta E\simeq  A_{N_0}(\Delta N)^2$. As shown in the inset, the coefficient $A_{N_0}$ values decrease approximately as $A_{N_0}\sim 1/N_0$, implying that the energy cost of changing the condensate charge vanishes in the thermodynamic limit \cite{wietek2017}. 
The collapse of the tower signals spontaneous breaking of the total $U(1)_c$ charge symmetry, identifying the strong-attraction phase as an interlayer-paired SF.

Having established the FQSH and paired-SF regimes, we next focus on the transition point. Here we used two complementary methods. For a critical point described by a CFT, the gaplessness condition is \cite{Cardy1984,Cardy1985}
\begin{equation}
  E_{n}-E_0
  =  \frac{v}{R}\Delta_{n}
  +o(R^{-1}),
  \label{eq:sphere_gap_scaling}
\end{equation}
where $R\propto\sqrt{N}$ is the typical length scale of the sphere, $v$ is the nonuniversal
velocity, and $\Delta_{n}$ is the scaling dimension of the
corresponding CFT operator. 
Figure~\ref{fig:scaled-gap-crossing}(a) displays the rescaled low-energy gaps
for $N=4$--$7$. Near the transition point, the low-lying excitations for different system sizes nearly collapse, 
whereas they separate systematically on either side. This behavior provides a common finite-size crossing region and locates the transition point.
Moreover, as a separate method, we introduce a spectral cost function to quantify the agreement with the XY CFT spectrum:
\begin{equation}
  f_{\mathrm{cost}}^{(3)}(v_0,v_1;N) =
  \sum_{\mathcal O\in
  \{\epsilon,\partial\epsilon,\Box\epsilon\}}
  \left|
    \Delta_{\mathcal O}^{(N)}(v_0,v_1)  -
    \Delta_{\mathcal O}^{\mathrm{XY}}
  \right|\,,
  \label{eq:spectral_cost}
\end{equation}
where $\Delta_\epsilon\approx 1.51136 $ is the dimension of leading scalar primary~\cite{Poland_2019,Chester_2020}, $\partial_\mu\epsilon$ is its spin-one first descendant, and $\Box\epsilon$ is its scalar second-level descendant. We define the finite-size scaling dimension by fixing the conserved-current dimension to $\Delta_J=2$~\cite{Zhu2023FuzzySphere}, hence
$\Delta_{n}^{(N)}=2(E_{n}-E_0)/(E_J-E_0)$,
where $E_0$ is the neutral ground-state energy and $E_J$ is the energy of the current operator $J_\mu$. 
This cost simultaneously tests the value of the leading scalar dimension and the integer-spaced descendant structure expected from conformal symmetry.
As shown in Fig.~\ref{fig:laughlin-fidelity}(b),
$f_{\mathrm{cost}}^{(3)}$ shows a pronounced minimum that occurs near $(v_0^*,v_1^*)=(-0.45,-0.05)$.
The location of this minimum agrees with the critical point inferred independently from the size collapse of energy gaps in
Fig.~\ref{fig:scaled-gap-crossing}(a). We therefore take
$(v_0,v_1)=(-0.45,-0.05)$ as the optimal critical point to analyze the operator spectrum.

\begin{figure}[tbp]
    \centering
    \includegraphics[width=\columnwidth]{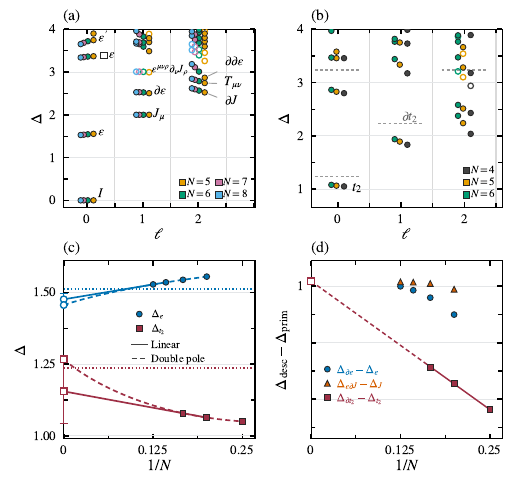}
    \caption{
    Critical operator spectra in the (a) $(Q_c,Q_s)=(0,0)$ and (b) $(-2,0)$ sectors, with low-lying XY$^*$ CFT operators labeled.
    Filled/open circles denote $\mathcal M_T=+1/-1$.
    (c) Finite-size dependence of scaling dimensions of the thermal scalar $\Delta_\epsilon$ and rank-2 operator
     $\Delta_{t_2}$.
    (d) Finite-size dependence of descendant spacings $\Delta_{\partial\epsilon}-\Delta_\epsilon$,
    $\Delta_{\varepsilon\partial J}-\Delta_J$, and
    $\Delta_{\partial t_2}-\Delta_{t_2}$,
    where $\varepsilon\partial J$ denotes
    $\varepsilon^{\mu\nu\rho}\partial_\nu J_\rho$.
    In (c), solid lines fit $\Delta_\infty+a/N$ to the two largest sizes;
    dashed lines fit $\Delta_\infty+a/N+b/N^{1.44}$ to all displayed sizes.
    Dotted horizontal lines indicate the theoretical predictions of XY$^*$ CFT. }
    \label{fig:critical-towers}
\end{figure}

The operator spectra at the critical point are shown in
Fig.~\ref{fig:critical-towers}(a)-(b). 
First, the leading scalar field is identified as $\epsilon$, which is the thermal scalar of the XY
CFT and represents the relevant deformation that tunes the system across the transition. Its scaling dimension drifts from $\Delta_\epsilon=1.555$ at $N=5$ to $1.528$ at $N=8$. The linear extrapolation gives $\Delta_\epsilon=1.476(20)$, compared with the theoretical value $1.511$ \cite{Poland_2019,Chester_2020}.
Within the resolved neutral spectrum, $\epsilon$ is the only nontrivial symmetry-preserving relevant scalar, consistent with the critical point reached by tuning a single microscopic parameter.

Second, the descendants $\partial_\mu \epsilon, \square \epsilon$ provide a particularly clear test of conformal organization. The spacing between $\epsilon$ and $\partial_\mu \epsilon$ evolves as $\Delta_{\partial\epsilon}-\Delta_\epsilon
  =  0.950,\ 0.980,\ 0.993,\ 0.999$ for $N=5,6,7,8$, and thus converges rapidly toward the unit spacing required by
conformal symmetry [Fig.~\ref{fig:critical-towers}(d)]. 
Additionally, the descendant $\square \epsilon$ lies between
$3.339$ and $3.369$, approximately $5\%$ below the XY prediction
$3.511$. The $\epsilon$ primary and its descendants therefore provide a robust finite-size signature of conformal symmetry.

Third, the conserved-current multiplet provides a further test of the emergent conformal symmetry. The level-one descendants of $J_\mu$ include a spin-$1$ operator $\epsilon^{\mu\nu\rho}\partial_\nu J_\rho$ and a spin-$2$ operator $\partial_{\mu}J_{\nu}$, both expected at $\Delta=3$. The spin-$1$ descendant is identified with the lowest state in the $(\ell,\mathcal{M}_T)=(1,-)$ sector, with scaling dimensions $2.995$, $3.005$, $3.007$, and $3.008$ for $N=5–8$ [Fig.~\ref{fig:critical-towers}(d)], respectively. The spin-$2$ descendant is likewise observed near $\Delta=3$ in the $(\ell,\mathcal{M}_T)=(2,+)$ sector. 
In addition, the deviations in large-spin sectors are attributed to the scattering from magnetorotons~\cite{GMP1986}, which can be paritally eliminated by using the projection method~\cite{sm}.

Last but not least, the rank-2 operator $t_2$ is found in the $(Q_c,Q_s)=(-2,0)$ sector in Fig.~\ref{fig:critical-towers}(b). It is directly related to the SF order parameter in the effective theory in Eq.~\eqref{eq:xy_star_action}. We estimate $\Delta_{t_2}=1.16(11)$ using the same fitting procedure, consistent with the value $\sim 1.236$ from conformal bootstrap \cite{Poland_2019,Chester_2020} and from Monte Carlo calculation \cite{Hasenbusch2011}. In particular, identification of this operator is crucial to understand the nature of fractionalized XY* criticality and its difference from an ordinary XY criticality.
The distinction lies in how the CFT operators are embedded into the
microscopic Hilbert space: the fundamental critical field $\phi\sim s\bar s$ is
a fractionalized semion--antisemion bound state, whereas the local interlayer-pair operator satisfies
$b_\uparrow b_\downarrow\sim\phi^2$.
Consequently, the physical SF order parameter corresponds to the charge-two primary of the XY CFT rather than to its
fundamental charge-one field. 

{\bf \em Summary and discussion.---}We have studied the interaction-driven transition between a FQSH state, with doubled-semion topological order, and an interlayer paired SF in an opposite-flux bosonic bilayer. Condensation of a bosonic semion--antisemion bound state destroys the topological order and spontaneously breaks the total-charge $U(1)$ symmetry. Using fuzzy-sphere spectroscopy, we resolved low-energy conformal families consistent with the 3D XY$^{*}$ Wilson-Fisher CFT. The fractionalized nature of the transition is reflected in the physical pair order parameter, which corresponds to the charge-two primary of the XY CFT rather than its fundamental field.

In End Matter, we present evidence for a $\mathbb{Z}_3$ counterpart of this transition in an \emph{electronic} quantum Hall bilayer at fillings $(\nu_\uparrow,\nu_\downarrow)=(1/3,2/3)$ with Coulomb interactions and finite layer width.
Here, the physical interlayer exciton corresponds to a cubic composite of the fractionalized critical field.
An independent analysis of its ground-state correlator yields the scaling dimension $\Delta_\chi=2.12(2)$, consistent with the charge-three XY primary and clearly distinct from the fundamental field of an ordinary XY transition. These results provide a concrete route to probing XY$^{*}$ criticality in existing semiconductor quantum Hall bilayer experimental setups~\cite{Spielman2001,Kellogg2002,Tutuc2003,Tutuc2004,Champagne2008}.

More broadly, our results suggest that multicomponent quantum Hall systems provide a versatile platform for investigating anyon condensation \cite{Ji2026,MengCheng2026}, symmetry breaking, and emergent conformal symmetry.
Beyond XY*, in the SM~\cite{sm} we show that introducing pair tunnelling in the bosonic model provides a route to an expected Ising* transition from time-reversal-symmetric doubled-semion topological order to a trivial phase, analogous to the $\mathbb{Z}_2$ transition between the toric code and a paramagnet\cite{Vidal2009,Tupitsyn2010,FCWu2012,Somoza2021,Schuler_2023,Oppenheim2024}.

    
{\bf \em Acknowledgements.---}W.Z. thanks Zheng Zhou for sharing the new version of FuzzifiED package \cite{zhou2025fuzzifiedjuliapackage} before publication. M.Y. and Z.P. would like to thank Joseph Taylor for useful discussions. L.D.H. and W.Z. are supported by NSFC under No. 12625407,12474144. L.D.H. is supported by the Fundamental Research Funds for the Central Universities. L.Z. is supported by the National University of Singapore start-up grants A-0009991-00-00 and A-0009991-01-00. Any opinions, findings and conclusions or recommendations expressed in this material are those of the auhtors and do not reflect the views of National University of Singapore. Numerical simulations have used the DiagHam~\cite{diagham} and FuzzifiED~\cite{FuzzifiED} libraries.  M.Y. and Z.P. acknowledge support by the Leverhulme Trust Research Leadership Award No. RL-2019-015 and EPSRC Grants EP/Z533634/1,  EP/W524372/1, UKRI1337, UKRI4851. This research was supported in part by grant NSF PHY-2309135 to the Kavli Institute for Theoretical Physics (KITP). Some of the numerical simulations were undertaken on the HPC systems at the University of Leeds, UK.


\bibliography{CFT_Sphere.bib}

\newpage
\cleardoublepage

\onecolumngrid
\begin{center}
{\large {\bf End Matter}}
\end{center}
\vspace{1em}

\twocolumngrid

{\bf \em XY$^{*}$ criticality in a fermionic bilayer.---}The bosonic model of the main text has a fermionic counterpart of direct relevance to GaAs quantum Hall experiments: a bilayer at total filling $\nu_T=1$ with imbalanced layer fillings $(\nu_\uparrow, \nu_\downarrow) = (1/3, 2/3)$~\cite{YahuiZhang2023}. Performing a particle-hole transformation of the $\downarrow$ layer maps it onto a $(1/3,-1/3)$ electron-hole bilayer, where the two layers effectively experience opposite magnetic fields. At large layer separation, the ground state is a pair of decoupled Laughlin states of opposite chirality -- the analog of the FQSH state considered in the main text, while at small separations it is the interlayer exciton condensate~\cite{annurevEisenstein}.  However, in the bilayer case there is no exact mapping between FQH states of bosons and fermions; moreover, the particle-hole transformation of $\downarrow$ is only strictly valid when the layers are decoupled. Nevertheless, we will show that this fermionic bilayer also realizes a transition in XY$^{*}$ universality class~\cite{YahuiZhang2023}, even in the presence of realistic Coulomb interactions. 

\begin{figure}[t]
  \centering
  \includegraphics[width=\linewidth]{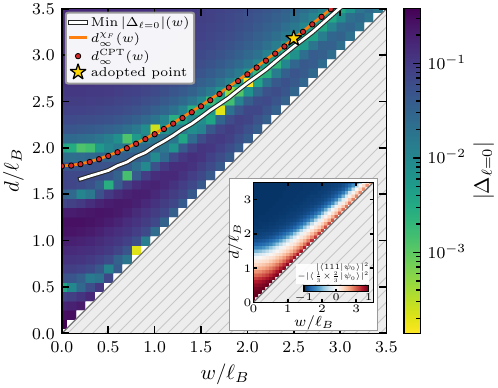}
  \caption{\label{fig:1323_111_phasediagram} Phase diagram of the fermionic $(1/3, 2/3)$ bilayer as a function of layer width $w$ and separation $d$. Color bar shows the magnitude of the extrapolated gap in $\ell = 0$ sector, $|\Delta_{\ell=0}|$. The gap is obtained by linear extrapolation in $1/\sqrt{N_e}$ from $N_e = 10$-$19$ data. The minimum of $|\Delta_{\ell=0}|$ at each $w$ is marked by the white line. For completeness, we also show the critical separation $d_\infty^{\chi_F}(w)$ from the extrapolated fidelity susceptibility peaks (orange line, for $N_e = 10$-$22$), as well $d_\infty^{\rm CPT}(w)$ obtained from conformal perturbation theory (red points), see the SM~\cite{sm} for details. The star marks the point $(w, d) = (2.5, 3.18)$ used for spectral analysis. The hatched region $d < w$ is unphysical in our model. Inset: the difference in overlap, $|\langle 111|\psi_0\rangle|^2 - |\langle 1/3 \times 2/3|\psi_0\rangle|^2$, between the exact ground state $|\psi_0\rangle$ and two trial states, the Halperin 111 state~\cite{Halperin1983} and the product of decoupled Laughlin states, at $N_e=16$. This shows the exciton condensate at small $d$  (red region) and the decoupled Laughlin phase at large $d$ (blue). }
\end{figure}

We place $N_e$ electrons on the fuzzy sphere with $N_\phi = N_e - 1$ flux quanta and the electron density fixed to $1/3$ in $\uparrow$ layer and $2/3$ in $\downarrow$ layer.  The electrons in the same layer interact via $V^{\uparrow\uparrow} = V^{\downarrow\downarrow} = 1/\sqrt{r^2 + w^2}$, the Coulomb interaction screened by finite layer width $w$~\cite{ZhangDasSarma}. Electrons in opposite layers interact via $V^{\uparrow\downarrow}(r) = 1/\sqrt{r^2 + d^2}$, where $d>w$ is the interlayer distance. Increasing $d$ drives the exciton condensate into the decoupled Laughlin phase along a critical
line $d_c(w)$, which we map out in Fig.~\ref{fig:1323_111_phasediagram}. Over the range of widths studied, the critical separation, $d_c/\ell_B \approx 1.8$--$3.5$, is consistent with previous experiments on exciton condensates in GaAs bilayers~\cite{Spielman2001,Kellogg2002,Tutuc2003,Tutuc2004,Champagne2008}. This estimate is also consistent between several metrics, including the minimum of the gap, overlap with trial states, the fidelity susceptibility and conformal perturbation theory, which are discussed in the SM~\cite{sm}.

{\bf \em Operator spectrum at criticality.---}The interlayer excitation of the decoupled Laughlin phase is a fractional exciton $\varphi$: a charge-$e/3$ quasiparticle in one layer bound to a
quasihole in the other. The transition is described by the $\mathrm{O(2)}$ Wilson--Fisher theory of $\varphi$ coupled to a $\mathbb{Z}_3$
gauge field (XY$^\ast$), the $\mathbb{Z}_3$ analogue of Eq.~(\ref{eq:xy_star_action})~\cite{YahuiZhang2023}, in which only operators with $\mathrm{O(2)}$ charge $q \equiv 0 \pmod 3$ are local. The neutral sector, and with it $\nu$ and all thermodynamic exponents, is identical to
that of O(2) and cannot reveal that the transition proceeds out of a fractionalized phase. That information is carried by the physical
interlayer order parameter $\chi \sim \varphi^3$, whose dimension $\Delta_\chi = 2.1086(3)$ replaces $\Delta_{\sigma_1} = 0.519$ of the standard
O(2)~\cite{Chester2020O2,Liu2020O2,Hasenbusch2026}, making it the key signature of XY$^\ast$. Measuring it requires resolving states by their charge $q$, which the bilayer provides directly: since the electron number of each layer is conserved, the Hamiltonian has a $\mathrm{U(1)}$ symmetry under which the interlayer exciton $\Phi^\dagger = c^\dagger_\uparrow c_\downarrow
\sim \chi^\dagger$ carries $Q = 1$. With $N_\uparrow = (N_e + 2 + q)/3$ electrons, the $\uparrow$ layer holds $q$ charge-$e/3$ quasiparticles on top
of the $1/3$ Laughlin state and the $\downarrow$ layer the opposite excitations, so each unit of $q$ is one $\varphi$ and $Q = q/3$. Sizes
$N_e \equiv 1 \pmod 3$ host the vacuum and the $q \in 3\mathbb{Z}$ sectors containing $\chi$, while the other, $\mathbb{Z}_3$-twisted sizes host
$\sigma_1$ ($q = \pm 1$) and $t_2$ ($q = \pm 2$), so each sector is selected by the choice of $N_e$ alone, without defects or extra flux.
Within a sector, states are labelled by angular momentum $\ell$ and by $\mathcal{PH}\circ\mathcal{Z}$ (particle--hole conjugation of both
layers combined with layer exchange), which plays the role of $M_T$ in the main text and acts as spatial parity on the CFT.

\begin{figure}[t]
  \centering
  \includegraphics[width=\linewidth]{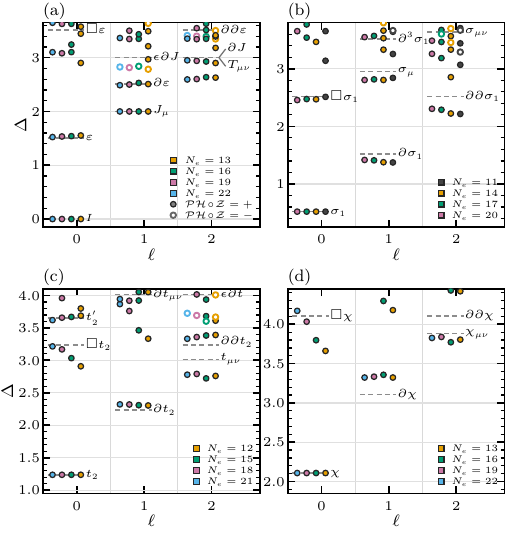}
  \caption{\label{fig:1323_111_coulombonly_spectra} Critical spectrum at $w = 2.5$, $d = 3.18$ for the four largest sizes $N_e$ (different colors) in different charge $q=0, -1, -2, -3$ sectors. The spectra are normalized by setting $\Delta_{J_\mu} = 2$.  Filled (open) symbols correspond to the $+$ ($-$) eigenvalue of $\mathcal{PH}\circ\mathcal{Z}$. Dashed lines are the expected O(2) scaling dimensions. (a) $q = 0$, after projecting out the $\ell=2$ magnetoroton state of each $\mathcal{PH}\circ\mathcal{Z}$ sector~\cite{sm}. 
(b) $\sigma_1$ ($q = -1$), (c) $t_2$ ($q = -2$) and (d) $\chi$ ($q = -3$), each with its lowest $\ell = 0$ state pinned to the XY$^\ast$ value. Here $\epsilon\partial J \equiv \epsilon^{\mu\nu\rho}\partial_\nu J_\rho$ and $\epsilon\partial t \equiv \epsilon^{\mu\nu\rho}\partial_\nu t_{\rho\lambda}$ denote curl descendants.}
\end{figure}

Figure~\ref{fig:1323_111_coulombonly_spectra} shows the spectrum at the optimal point ($w=2.5$, $d = 3.18$) in the sectors $q = 0, -1, -2, -3$, for the four largest sizes of each. We normalize the spectrum by setting $\Delta_{J_\mu} = 2$. For clarity, in the neutral sector [panel (a)] the long-wavelength magnetoroton of the two Laughlin states is projected out, as described in the SM~\cite{sm}. The structure of the vacuum sector in Fig.~\ref{fig:1323_111_coulombonly_spectra}(a) is consistent with XY$^{*}$ CFT. For example, at system size $N_e = 22$, we obtain $\Delta_\varepsilon = 1.524$ (against the expected 1.511), $\Delta_{\partial\varepsilon} = 2.488$ (vs. 2.511), $\epsilon^{\mu\nu\rho}\partial_\nu J_\rho$
is at $2.83$ (vs. 3) and $\varepsilon'$ at $3.65$ (vs. 3.794). Higher up in the spectrum, the agreement becomes poorer: the lowest natural $\ell = 2$ state $T_{\mu\nu}$ lies at $2.59$ instead of 3, an extra $\ell = 0$ level appears at $3.10$, below $\Box\varepsilon$, and a few non-roton states remain at $\ell = 3, 4$ below $\Delta = 3.6$. Nevertheless, in the SM~\cite{sm} we show that a small pseudopotential deformation of the Coulomb interaction removes most of these deviations, supporting the XY$^{*}$ critical point.

Moving charge between the layers adds a non-CFT energy with parts linear and even in $Q$. Averaging the $\pm Q$ sectors removes the linear part,
but the even part remains and cannot be removed reliably. The remaining non-CFT energy is also large, e.g., the $\pm Q$-averaged exciton gap exceeds $\Delta_\chi$ by $\approx 2$, and it does not decrease over the available $N_e$ range, so it cannot be extrapolated away. In Fig.~\ref{fig:1323_111_coulombonly_spectra}(b)-(d) we therefore pin the lowest natural $\ell = 0$ state of each charged
sector to its XY$^\ast$ primary: $\sigma_1$ ($q = -1$, $0.519$), $t_2$ ($q = -2$, $1.236$) and $\chi$ ($q = -3$, $2.109$). We use the vacuum velocity,
$\alpha = 2/(E_J - E_0)$. Twisted sizes ($N _e\not\equiv 1 \bmod 3$) have no vacuum, so there $\alpha$ is interpolated linearly in $1/N_e$ between the neighbouring sizes $N_e \equiv 1 \pmod 3$. Alternatively, interpolating the velocity $v \propto R/\alpha$ instead changes the spacings by at most $0.4\%$. Using this procedure, we can identify in each sector the lowest descendants and the next primaries ($\sigma_\mu$, $t_{\mu\nu}$, $\chi_{\mu\nu}$).

\begin{figure}[t]
  \centering
  \includegraphics[width=\linewidth]{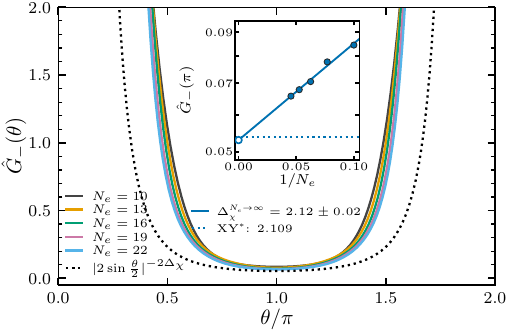}
  \caption{\label{fig:1323_111_coulombonly_exciton} Exciton correlator $\hat G_-(\theta)$ in Eq.~(\ref{eq:normalizedG}), computed at $w = 2.5$, $d = 3.18$ for system sizes $N_e = 10$-$22$. Dotted line is the primary form $|2\sin(\theta/2)|^{-2\Delta_\chi}$ with the XY$^\ast$ value
$\Delta_\chi = 2.109$. Inset: $\hat G_-(\pi)$ against $1/N_e$ (log scale) with the fit $\ln\hat G_-(\pi) = -2\Delta_\infty\ln 2 + B/N_e$, which gives
$\Delta_\infty = 2.12(2)$ (open circle). The dotted line marks $2^{-2\Delta_\chi}$ for XY$^\ast$.}
\end{figure}

{\bf \em Exciton correlator.---}Since $\Delta_\chi$ is pinned in the charged spectrum, we measure it independently from the ground-state correlator
$G(\theta) = \langle \Phi^\dagger(\theta)\Phi(0)\rangle$ of the lattice exciton $\Phi^\dagger = c^\dagger_\uparrow c_\downarrow$, which involves no charging energy. We compute the normalized correlator,
\begin{equation}\label{eq:normalizedG}
 \hat G_-(\theta) = \langle \Phi^\dagger(\theta)\Phi(0)\rangle /|\langle\chi|\Phi_{0,0}|0\rangle|^2\,,  
\end{equation}
dividing it by the weight  of the lowest $Q = -1$,
$\ell = 0$ state, which removes the non-universal overall factor. If $\Phi$ were a pure primary, the normalized correlator would decay according to 
$\hat G_-(\theta) = |2\sin(\theta/2)|^{-2\Delta_\chi}$, which we test in  Fig.~\ref{fig:1323_111_coulombonly_exciton}. 

Lattice corrections are
smallest at $\theta = \pi$, where the two operator insertions are furthest apart, so we use $\hat G_-(\pi) = 2^{-2\Delta_\chi}$. The descendants of $\chi$ in the lattice operator give corrections $\propto 1/N_e$. The fit $\ln\hat G_-(\pi) = -2\Delta_\infty\ln 2 + B/N_e$
gives $\Delta_\infty = 2.12 \pm 0.02$ (inset), in agreement with XY$^\ast$ value of 2.109 and far from the XY value, 0.519. Adding the subleading correction from the next charge-3 primary $\chi_{\mu\nu}$ gives $2.18 \pm 0.09$, consistent with the leading fit but not a significant improvement.

As pointed out in Ref.~\cite{YahuiZhang2023}, $\Delta_\chi$ has a direct experimental signature. Since $G(\theta)$ is the equal-time counterpart of the interlayer tunnelling
response, large $\Delta_\chi$ suppresses the critical tunnelling conductance at low interlayer bias, whereas at an
ordinary XY transition, with $\Delta_{\sigma_1} \approx 0.5$, it is strongly enhanced.

\clearpage
\onecolumngrid
\appendix

\begin{center}
\textbf{\large Supplementary Material for ``
Fractionalized XY$^*$ criticality in bilayers with opposite magnetic flux" }\\[5pt]
\vspace{0.1cm}
\begin{quote}
{\small 
In this Supplementary Material, we present a detailed introduction to the XY$^*$ field theory from a different perspective, details of our microscopic model and its realization on the fuzzy sphere, discussing in detail the symmetry breaking in the superfluid phase. We also present further numerical data supporting the main text, including the magnetoroton analysis and the identification of the critical point in the fermionic bilayer model. Finally, we also demonstrate that our approach can be generalized to the Ising$^*$ transition in a similar setup. 
} \\[20pt]
\end{quote}
\end{center}
\setcounter{figure}{0}
\renewcommand{\thefigure}{S\arabic{figure}}

\section{S1: XY$^*$ theory from gauging perspective}

To make the distinction between our realization of the XY$^{*}$ theory and that of Ref.~\cite{Isakov_2012} more concrete, it is useful to look at the ungauged versions of the two setups. 
The ungauged version of the setup in Ref. \cite{Isakov_2012} can simply be taken as a bosonic system with a $U(1)$ symmetry that undergoes an ordinary insulator-superfluid transition \cite{PhysRevB.40.546}, under which the elementary boson carries charge-$1/2$ using the convention therein. To see that gauging the $\mathbb{Z}_2$ subgroup of this $U(1)$ symmetry results in the setup in Ref. \cite{Isakov_2012}, we start from the insulator side. A useful way to characterize the insulator for our purpose is via its edge theory, which has the effective Lagrangian \cite{Levin2012}
\begin{equation}
    \mathcal{L}=\frac{1}{2\pi}\partial_t\phi_1\partial_x\phi_2+\cdots,
\end{equation}
where $\cdots$ represents non-topological term. The $U(1)$ symmetry acts by
\begin{equation}
    \phi_1\rightarrow\phi_1,
    \quad
    \phi_2\rightarrow\phi_2+\theta/2,
\end{equation}
with $\theta$ being the $U(1)$ phase factor and the $1/2$ factor capturing the charge of the boson. The Kac-Moody algebra dictated by the above Lagrangian implies that the $\mathbb{Z}_2$ flux is $e^{i\phi_{\rm flux}}=e^{i\phi_1/2}$, and gauging $\mathbb{Z}_2$ amounts to writing the Lagrangian in terms of $\phi_{\rm flux}$ and $\phi_2$:
\begin{equation}
    \mathcal{L}=\frac{1}{\pi}\partial_t\phi_{\rm flux}\partial_x\phi_2+\cdots,
\end{equation}
which is precisely the edge theory of a $\mathbb{Z}_2$ quantum spin liquid, in agreement with Ref.~\cite{Isakov_2012}. Moreover, after gauging, only $e^{2i\phi_{\rm flux}}$ and $e^{2i\phi_2}$ are local operators, so all local operators only carry integer charge now and there is no charge-$1/2$ local excitation anymore. Consequently, the ungauged charge-$1/2$ superfluid after gauging becomes a charge-1 superfluid, and the transition described by the standard XY theory becomes the XY* theory, where all operators in the XY theory with half charge are no longer local.

In contrast, the ungauged version of our setup is a bosonic system with a $U(1)\times\mathbb{Z}_2$ symmetry that undergoes a transition between a nontrivial $\mathbb{Z}_2$ SPT phase and a superfluid. More precisely, the edge theory of the SPT phase is \cite{Levin2012a}
\begin{equation}
    \mathcal{L}=\frac{1}{2\pi}\partial_t\phi_1\partial_x\phi_2+\frac{1}{2\pi}\partial_t\phi_3\partial_x\phi_4+\cdots,
\end{equation}
with the $U(1)\times\mathbb{Z}_2$ acting by
\begin{equation}
    \begin{split}
    &U(1): \phi_{1, 3}\rightarrow\phi_{1, 3},
    \quad
    \phi_{2, 4}\rightarrow\phi_{2, 4}+\theta,\\
    &\mathbb{Z}_2: \phi_{1, 2}\rightarrow\phi_{1, 2}+\pi,
    \quad
    \phi_{3, 4}\rightarrow\phi_{3, 4}.
    \end{split}
\end{equation}
Here $\phi_{1, 2}$ can be viewed as the ungauged version of $s\bar s$ that will be condensed, $\phi_{3, 4}$ are the original local boson, and all these fields carry integer charges under $U(1)$. The Kac-Moody algebra dictates that now the $\mathbb{Z}_2$ flux is $e^{i\phi_{\rm flux}^{(\pm)}}=e^{i\frac{\phi_1\pm \phi_2}{2}}$, and gauging $\mathbb{Z}_2$ amounts to write the Lagrangian in terms of $\phi_{\rm flux}^{(\pm)}$ and $\phi_{3, 4}$:
\begin{equation}
    \mathcal{L}=\frac{1}{2\pi}(\partial_t\phi_{\rm flux}^{(+)}\partial_x\phi_{\rm flux}^{(+)}-\partial_t\phi_{\rm flux}^{(-)}\partial_x\phi_{\rm flux}^{(-)})+\frac{1}{2\pi}\partial_t\phi_3\partial_x\phi_4+\cdots,
\end{equation}
which is precisely the edge theory of a double-semion phase. In particular, there are still local fields (i.e., $e^{i\phi_4}$) carrying charge-$1$. The ungauged charge-$1$ superfluid becomes a paired superfluid after gauging (but there are still gapped charge-$1$ excitations, which are the original bosons), and the transition is precisely described by our XY* theory.

\section{S2: Lowest-Landau-level matrix elements and symmetry sectors}
\label{app:matrix-elements}

\subsection{Orbitals in opposite magnetic fields}

The LLL at monopole charge $q=\pm s$ contains $2s+1$ orbitals,
with angular-momentum quantum numbers $(s,m)$ and $-s\leqslant m\leqslant s$.
We normalize the monopole harmonics as 
$\int d\boldsymbol{\Omega}\,|Y_{s,m}^{(q)}(\boldsymbol{\Omega})|^2=1$.
The phase convention relating the two layers is
\begin{equation}
    Y_{s,m}^{(-s)}(\boldsymbol{\Omega})
    =(-1)^{s-m}\bigl[Y_{s,-m}^{(s)}(\boldsymbol{\Omega})\bigr]^*.
    \label{eq:app-opposite-orbitals}
\end{equation}
Thus reversing the field conjugates the orbital wave function and
reverses its magnetic quantum number. Both sets of orbitals carry
angular momentum $s$, so the generators acting on the projected
fields have the same form in the two layers:
\begin{equation}
L_z=\sum_{\sigma,m}m\,b^\dagger_{m,\sigma}b_{m,\sigma},
    \qquad
    L_+=\sum_{\sigma}\sum_{m=-s}^{s-1}
    \sqrt{(s-m)(s+m+1)}\,
    b^\dagger_{m+1,\sigma}b_{m,\sigma},
    \qquad L_-=L_+^\dagger.
    \label{eq:app-angular-momentum}
\end{equation}
The opposite magnetic fields enter through the orbital wave functions
in the interaction matrix elements.

\subsection{Interaction matrix elements}

To construct the interlayer matrix elements for opposite magnetic
fields, we first introduce a same-field interaction kernel. For a
rotationally invariant potential $U$, projection onto charge-$s$
orbitals gives
\begin{align}
    A[U]_{m_1m_2m_3m_4}
    &\equiv\int d\boldsymbol{\Omega}_1\int d\boldsymbol{\Omega}_2\,
    \bigl[Y_{s,m_1}^{(s)}(\boldsymbol{\Omega}_1)\bigr]^*
    \bigl[Y_{s,m_2}^{(s)}(\boldsymbol{\Omega}_2)\bigr]^*
    U(\boldsymbol{\Omega}_1,\boldsymbol{\Omega}_2)
    Y_{s,m_3}^{(s)}(\boldsymbol{\Omega}_2)Y_{s,m_4}^{(s)}(\boldsymbol{\Omega}_1)
    \notag\\
    &=\delta_{m_1+m_2,m_3+m_4}
    \sum_{J=0}^{2s}V_{2s-J}(2J+1)
    \begin{pmatrix}s&s&J\\m_1&m_2&-m_1-m_2\end{pmatrix}
    \begin{pmatrix}s&s&J\\m_4&m_3&-m_3-m_4\end{pmatrix}.
    \label{eq:app-same-field-kernel}
\end{align}
Here $J$ is the angular momentum of a same-field pair and
$V_r$ is the Haldane pseudopotential with relative angular momentum
$r=2s-J$. The two Wigner $3j$ symbols arise by resolving the pair
into total-angular-momentum eigenstates. The order $(m_4,m_3)$ in
the second symbol follows the annihilation-operator order used below.
We also denote this kernel by $A[V]$ when specifying its
pseudopotential coefficients. For identical bosons within a layer,
only even $r$ contribute: odd-$r$ pair states are antisymmetric
under exchange and are excluded by Bose statistics. In particular,
$V_1^{\mathrm{intra}}$ can be chosen arbitrarily without changing
$H_{\mathrm{intra}}$.
Equation~\eqref{eq:app-opposite-orbitals} gives the same intralayer
kernel for the down layer: the phases from conjugation and from
reversing all magnetic quantum numbers cancel.

For particles in opposite fields, we substitute
Eq.~\eqref{eq:app-opposite-orbitals} into the down-layer orbitals.
The resulting interlayer kernel is
\begin{align}
    B[V]_{m_1m_2m_3m_4}
    &=(-1)^{2s-m_2-m_3}
      A[V]_{m_1,-m_3,-m_2,m_4}
    \notag\\
    &=\delta_{m_1+m_2,m_3+m_4}
    (-1)^{2s-m_2-m_3}
    \sum_{J=0}^{2s}V_{2s-J}(2J+1)
    \begin{pmatrix}s&s&J\\m_1&-m_3&m_3-m_1\end{pmatrix}
    \begin{pmatrix}s&s&J\\m_4&-m_2&m_2-m_4\end{pmatrix}.
    \label{eq:app-opposite-field-kernel}
\end{align}
Conjugation interchanges the down-layer creation and annihilation
indices inside the same-field kernel. The selection rule remains
$m_1+m_2=m_3+m_4$, as required by conservation of $L_z$.
The coefficients $V_r^{\mathrm{inter}}$ therefore specify the
same-field kernel before this transformation; the opposite-field
matrix elements are given by Eq.~\eqref{eq:app-opposite-field-kernel}.
In this construction both $r=0$ and $r=1$ contribute to the
interlayer interaction.

In the ordered-pair normalization used for the numerical spectra,
the second-quantized Hamiltonian reads
\begin{align}
    H_{\mathrm{intra}}
    &=\sum_{\sigma=\uparrow,\downarrow}\sum_{m_1,m_2,m_3,m_4}
    A[V^{\mathrm{intra}}]_{m_1m_2m_3m_4}
    b^\dagger_{m_1,\sigma}b^\dagger_{m_2,\sigma}
    b_{m_3,\sigma}b_{m_4,\sigma},
    \label{eq:app-h-intra}\\
    H_{\mathrm{inter}}
    &=2\sum_{m_1,m_2,m_3,m_4}
    B[V^{\mathrm{inter}}]_{m_1m_2m_3m_4}
    b^\dagger_{m_1,\uparrow}b^\dagger_{m_2,\downarrow}
    b_{m_3,\downarrow}b_{m_4,\uparrow}.
    \label{eq:app-h-inter}
\end{align}
The factor of two in the interlayer term accounts for the two
ordered layer pairs. We set $V_r^{\mathrm{intra}}=\delta_{r,0}$
and $V_r^{\mathrm{inter}}=v_0\delta_{r,0}+v_1\delta_{r,1}$.
These expressions give real matrix elements and preserve spatial
rotations. Their many-body matrix elements follow by applying the
bosonic operators to normalized occupation states, including the
usual occupation-number square roots.

\subsection{Symmetry-adapted diagonalization}

We construct the occupation basis at fixed particle number in each
layer and fixed $L_z$,
\begin{equation}
    |\boldsymbol n\rangle
    =\prod_{\sigma,m}\frac{(b^\dagger_{m,\sigma})^{n_{m,\sigma}}}
    {\sqrt{n_{m,\sigma}!}}|0\rangle,
    \qquad
    \sum_m n_{m,\uparrow}=\sum_m n_{m,\downarrow}=N,
    \qquad \sum_{\sigma,m}m\,n_{m,\sigma}=0.
    \label{eq:app-occupation-basis}
\end{equation}
Working at $L_z=0$ retains one state from each integer-angular-momentum
multiplet. We determine $\ell$ from $L^2=(L_+L_-+L_-L_+)/2+L_z^2$
within a degenerate energy subspace we diagonalize $L^2$ to resolve
the multiplets.

Time reversal exchanges the layers and reverses $m$. With $K$
denoting complex conjugation in the occupation basis, its action
can be written as
\begin{equation}
    \mathcal M_T=P_TK,\qquad
    P_T b^\dagger_{m,\uparrow}P_T^{-1}
      =(-1)^{s-m}b^\dagger_{-m,\downarrow},\qquad
    P_T b^\dagger_{m,\downarrow}P_T^{-1}
      =(-1)^{s-m}b^\dagger_{-m,\uparrow}.
    \label{eq:app-time-reversal}
\end{equation}
The rotation about the $y$ axis acts on the orbitals as
\begin{equation}
    P_y b^\dagger_{m,\sigma}P_y^{-1}
      =(-1)^{s-m}b^\dagger_{-m,\sigma}.
    \label{eq:app-y-rotation}
\end{equation}
It maps $L_z$ to $-L_z$ and hence acts within the selected sector
when $L_z=0$. On a total-angular-momentum state it gives
$P_y|\ell,0\rangle=(-1)^\ell|\ell,0\rangle$, yielding the
quantum number $\mathcal R_y=(-1)^\ell$ used in the main text.

In the balanced sectors considered here, $s$ and $N$ are integers
and the combined orbital phase in either operation is
$(-1)^{2Ns-L_z}=1$. The operations therefore reduce to permutations
of occupation states: $P_y$ reflects the orbitals within each
layer, while $P_T$ also exchanges the layers. Since the Hamiltonian
is real, we use these commuting real matrices to resolve its
$\mathbb Z_2$ sectors. The time-reversal label $\mathcal M_T=\pm1$
refers to this real representation. A basis state in the block
$(\mathcal M_T,\mathcal R_y)=(\eta_T,\eta_y)$ is obtained from
\begin{equation}
    |\boldsymbol n;\eta_T,\eta_y\rangle
    \propto (1+\eta_TP_T)(1+\eta_yP_y)|\boldsymbol n\rangle.
    \label{eq:app-symmetry-projector}
\end{equation}
We retain one normalized nonzero vector from each symmetry orbit.
This construction reduces the Hilbert-space dimension and gives
the symmetry labels used to organize the low-energy levels and
conformal towers.

\section{S3: Anderson tower and the instability of the superfluid}
\label{app:anderson-instability}

The charge-sector spectrum measures both the superfluid rotor energy
and the stability against density fluctuations. We denote by
$E_s(N)$ the lowest energy with $N_\uparrow=N_\downarrow=N$ at a
fixed monopole strength $s$. For a reference sector $N_0$, we choose
$2s=2(N_0-1)$ and vary $N$ while keeping this flux fixed.
The displacement $\Delta N=N-N_0=Q_c/2$ therefore changes the number of
bosons in each layer within the same orbital Hilbert space, with $Q_s=0$.
Removing the term linear in $\Delta N$ gives
\begin{equation}
    \Delta E_s(\Delta N)=E_s(N_0+\Delta N)-E_s(N_0)
       -\mu_{\mathrm p}\Delta N
       \simeq A_{N_0}(\Delta N)^2,
    \label{eq:app-rotor-subtraction}
\end{equation}
where $\mu_{\mathrm p}$ is the chemical potential for adding one
boson to each layer. For the spectra below, we use
$\mu_{\mathrm p}=[E_s(N_0+1)-E_s(N_0-1)]/2$ and fit the corrected
energies to $A_{N_0}(\Delta N)^2$ with the curve fixed to pass
through the origin. The curvature can also be obtained locally
from the neighboring charge sectors,
\begin{equation}
    A_{N_0}^{\mathrm{loc}}
    =\frac{E_s(N_0+1)+E_s(N_0-1)-2E_s(N_0)}{2}.
    \label{eq:app-discrete-curvature}
\end{equation}
The linear subtraction does not affect this second difference.
For a stable superfluid, $A_{N_0}>0$ and decreases inversely with
system area. At fixed filling the area grows with $N_0$, giving
the $1/N_0$ behavior shown in the inset of Fig.~\ref{fig:laughlin-fidelity}(c).

To relate this spectrum to thermodynamic stability, let
$\mathcal A=4\pi R^2$ and $N_{\mathrm{tot}}=2N$.
The compressibility along the balanced-density direction obeys
\begin{equation}
    \kappa^{-1}
    =\frac{N_{\mathrm{tot}}^2}{\mathcal A}
       \frac{\partial^2 E_s}{\partial N_{\mathrm{tot}}^2}
    \simeq\frac{2N_0^2}{\mathcal A}A_{N_0}.
    \label{eq:app-compressibility}
\end{equation}
An upward-curving tower thus has positive compressibility.
When the curvature vanishes, the charge sectors become degenerate
after the chemical-potential subtraction and the compressibility
diverges. A downward-curving spectrum gives negative
compressibility: transferring particles from a lower-density
region to a higher-density region lowers the energy. This density
instability accounts for the unstable region on the strongly
attractive side of the superfluid regime.

\begin{figure}[tbp]
    \centering
    \includegraphics[width=\columnwidth]{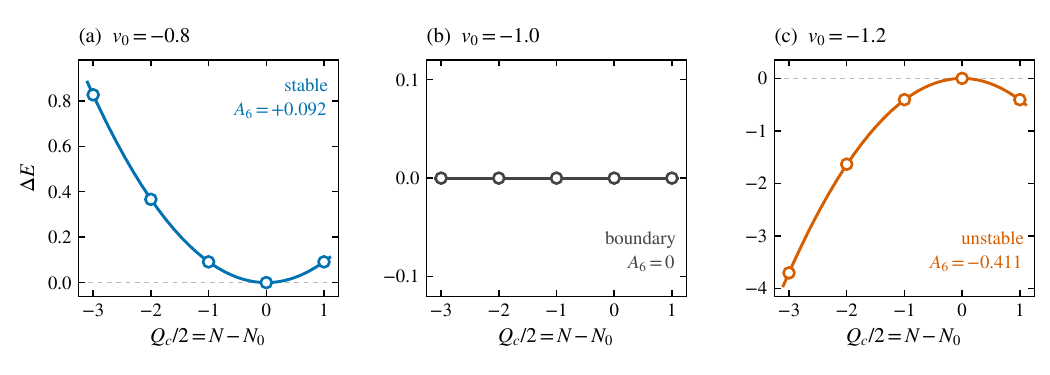}
    \caption{Charge-sector spectra across the superfluid stability
    boundary at fixed $v_1=-0.05$ and $N_\phi=10$.
    The reference sector has $N_0=6$ bosons per layer;
    the displayed sectors have $N=3,4,5,6,7$ and $Q_c/2=N-N_0$.
    Symbols show the sector ground-state energies after the
    chemical-potential subtraction in
    Eq.~\eqref{eq:app-rotor-subtraction}; curves are quadratic
    fits through the origin.
    (a) $v_0=-0.8$: positive curvature.
    (b) $v_0=-1.0$: vanishing curvature.
    (c) $v_0=-1.2$: negative curvature.
    Each panel uses its own vertical scale.
    The change of sign diagnoses the loss of positive
    compressibility on the strongly attractive side.}
    \label{fig:anderson-instability}
\end{figure}

Fig.~\ref{fig:anderson-instability} compares the charge-sector
spectra at $v_0=-0.8,-1.0,-1.2$ along the cut $v_1=-0.05$.
All three calculations use $N_\phi=10$, with reference particle
number $N_0=6$ in each layer and charge sectors $N=3,4,5,6,7$.
At $v_0=-0.8$, the levels form an upward-curving Anderson tower,
with $A_6\simeq0.092$. At $v_0=-1.0$, subtracting the linear
chemical-potential term leaves a flat sequence of charge-sector
energies. At $v_0=-1.2$, the curvature becomes negative,
$A_6\simeq-0.411$, and the balanced homogeneous state is unstable
against density fluctuations.

\section{S4: Spin and charge gaps, order parameter}

\subsection{Spin gaps}

\begin{figure}[ht]
    \centering
    \begin{minipage}{0.3\textwidth}
        \raggedright
        \textbf{(a)} $v_0=-0.8$ SF\par\vspace{2pt}
        \includegraphics[width=\linewidth]{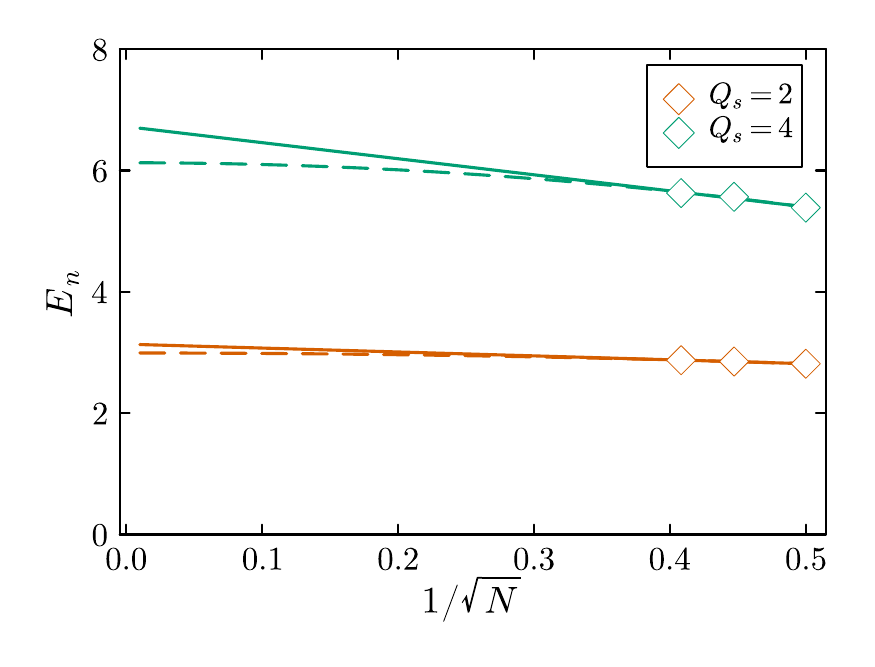}
    \end{minipage}
    \begin{minipage}{0.3\textwidth}
        \raggedright
        \textbf{(b)} $v_0=-0.45$ crit.\par\vspace{2pt}
        \includegraphics[width=\linewidth]{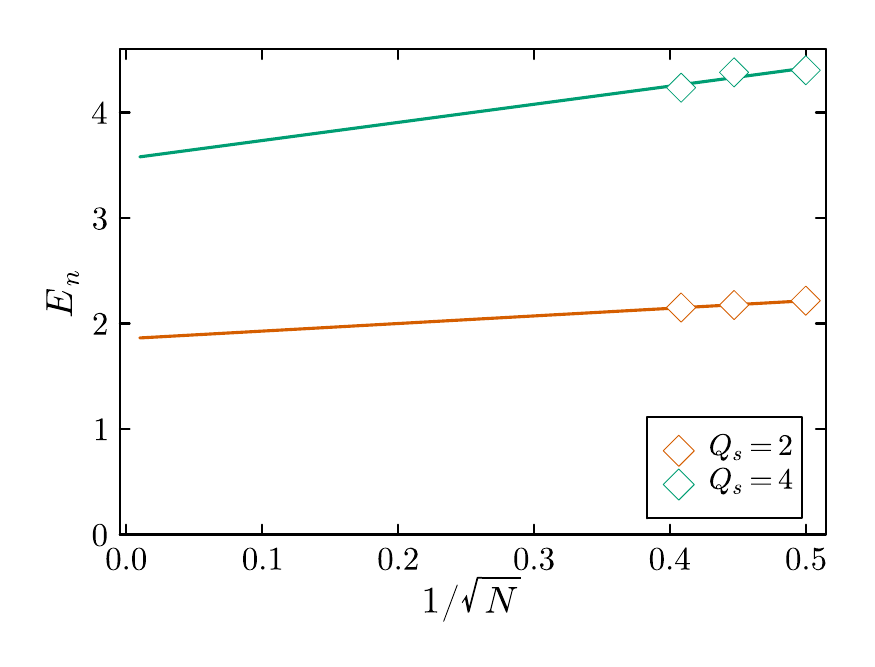}
    \end{minipage}
    \begin{minipage}{0.3\textwidth}
        \raggedright
        \textbf{(c)} $v_0=-0.1$ FQSH\par\vspace{2pt}
        \includegraphics[width=\linewidth]{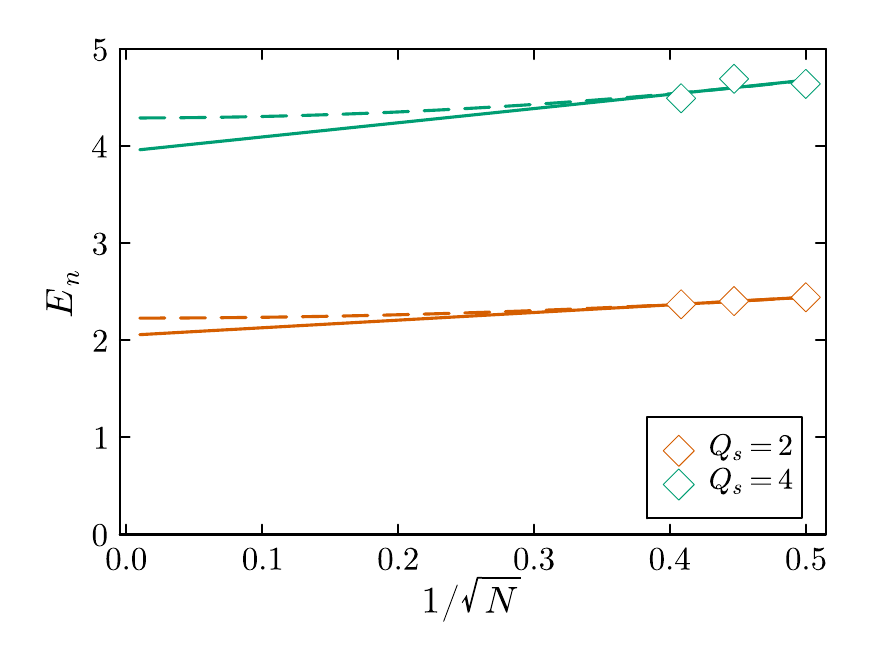}
    \end{minipage}
    \caption{The $Q_s=2,4$ gaps from finite size extrapolation in (a) the superfulid phase, (b) at XY$^*$ critical point, (c) in the FQSH phase. The solid lines use $E=a+b/\sqrt{N}$, the dashed lines use $E=a+b/N$.}
    \label{fig:spin-gaps}
    
\end{figure}

The continuous global symmetry group of our model can be written as $U(1)_c\times U(1)_s$, where $U(1)_c$ corresponds to the conservation of the total boson number in the bilayer, and $U(1)_s$ corresponds to the difference of the boson numbers on the two layers. Because of the charge of the scalar field $\phi(Q_c, Q_s)=(1, 0)$, the $U(1)_c$ symmetry is expected to be related to the $U(1)$ symmetry of the XY* critical point, while sectors with $Q_s\neq 0$ remain gapped at the critical point. Figure~\ref{fig:spin-gaps} confirms this behavior: we find the $U(1)_s$ charged sectors are gapped along the transition, which is consistent with the critical theory being neutral under this symmetry.

\subsection{Charge gap}
\label{app:charge-gap}
\raggedbottom

\begin{figure}[!ht]
    \centering
    \includegraphics[width=\textwidth]{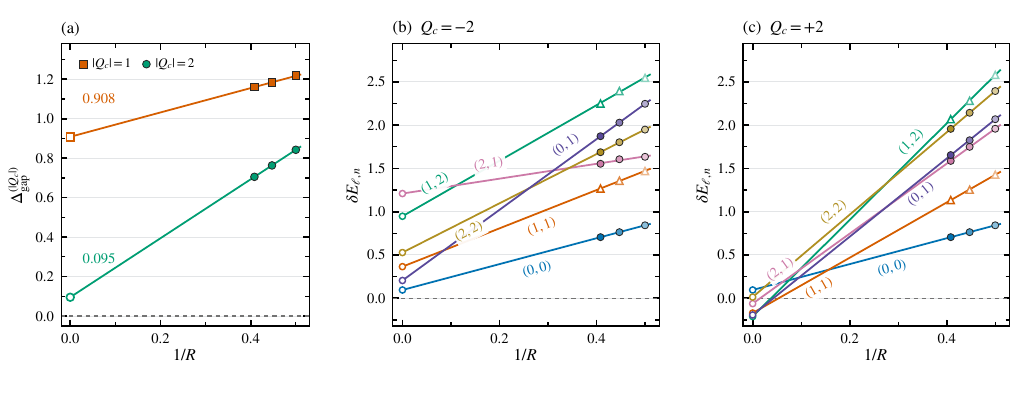}
    \caption{Charge gaps at $(v_0,v_1)=(-0.45,-0.05)$ for
    $N=4,5,6$, with $R=\sqrt N$.
    (a) Averaged $|Q_c|=1,2$ gaps.
    (b-c) Low-lying levels for $Q_c=-2$ and $+2$.
    Shades darken with increasing $N$.
    Circles/triangles denote layer-exchange parity $+1/-1$,
    and filled/open data symbols denote $\mathcal R_y=+1/-1$.
    The $\ell=2$ levels have $(\mathcal M_T,\mathcal R_y)=(+,+)$.}
    \label{fig:charge-gap}
\end{figure}

At the critical point, we study the total physical charge $Q_c$ associated with
the $U(1)_c$ symmetry introduced above.
As in the Anderson-tower analysis in the previous section, energies in different
charge sectors contain a chemical-potential contribution.
Writing $E_0(Q_c)$ for the charged ground-state energy and
$E_{00}=E_0(0)$, we choose
$\mu=[E_0(+2)-E_0(-2)]/4$ and subtract the linear term:
$\delta E_{\ell,n}(Q_c)=E_{\ell,n}(Q_c)-E_{00}-\mu Q_c$.
The particle-hole average cancels this term and gives
\begin{equation}
    \Delta_{\mathrm{gap}}^{(|Q_c|)}(N)=\frac{E_0(+|Q_c|)+E_0(-|Q_c|)}{2}-E_{00},
    \qquad |Q_c|=1,2.
    \label{eq:app-charge-gap}
\end{equation}
Figure~\ref{fig:charge-gap}(a) gives $\Delta_{\mathrm{gap}}^{(1)}(\infty)=0.908$ and
$\Delta_{\mathrm{gap}}^{(2)}(\infty)=0.095$ from linear fits in $1/R$.
Figure~\ref{fig:charge-gap}(b-c) shows the gap extrapolations for $Q_c=\pm 2$.
The finite $|Q_c|=1$ gap and small $|Q_c|=2$ gap
support a gapped single-particle channel and a gapless pair
channel. Together with the neutral XY spectrum, these are
the expected features of the XY$^*$ critical point, where
the local pair operator $b_\uparrow b_\downarrow\sim\phi^2$
is built from the fractionalized field $\phi$.

\subsection{Order parameter}

\begin{figure}[ht]
    \centering
    \begin{minipage}{0.5\textwidth}
        \includegraphics[width=\linewidth]{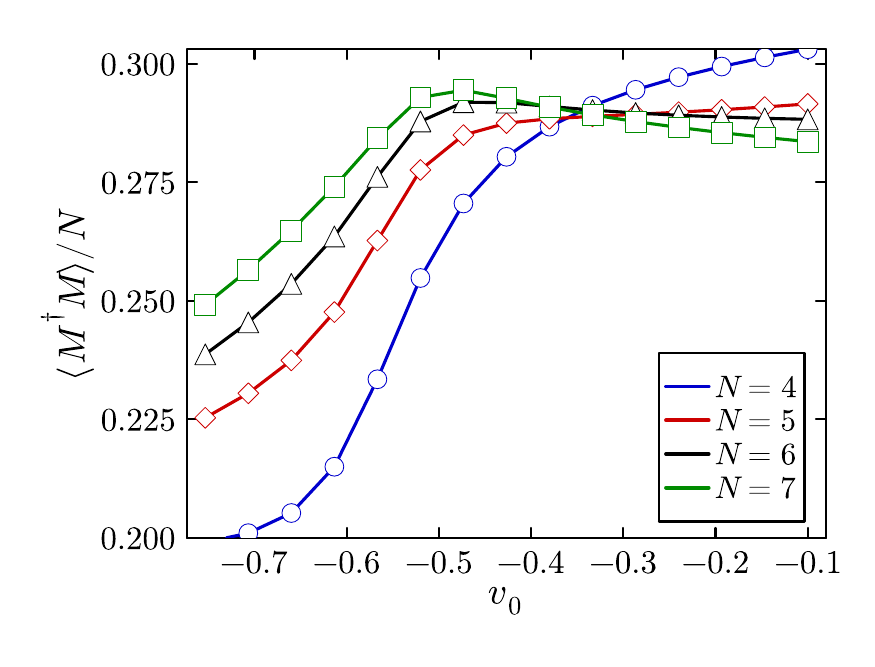}
    \end{minipage}
    
    \caption{ Order parameter $\langle M^\dagger M\rangle/N$ as a function of $v_0$, with $v_1=-0.05$ which is chosen as the critical value, for $N=4,5,6,7$.  }
    \label{fig:order-parameter}
    
\end{figure}

We define the order parameter as:
\begin{equation}
   M = \int d^2 r \, b_\downarrow (r)b_\uparrow (r),
\end{equation}
or written in the LLL-projected form:
\begin{equation}
   M = \sum_{m}b_{\uparrow m}b_{\downarrow-m}\left(-1\right)^{s+m}.
\end{equation}
$M$ has global charge $(Q_c,Q_s)=(-2,0)$, as determined by the symmetry, $ b_\downarrow (r)b_\uparrow (r)= c_t t(x)$, where $t(x)$ is the lowest scalar operator in the $Q_c=-2$ sector of XY fixed point. In the numerics, we compute the quantity $\langle M^\dagger M\rangle$, whose scaling law at the critical point is given by: 
\begin{equation}
\begin{split}
    \langle M^\dagger M\rangle&=\int d^2 x d^2 y |c_t|^2\langle t^\dagger(x)t(y)\rangle\\
    &=R^2 |c_t|^2 \int d^2 r\frac{1}{|r|^{2\Delta_t}}\\
    &=|c_{t}|^{2}\pi R^{2}\frac{1}{2\Delta_{t}-2}\left\{ \frac{1}{\epsilon^{2\Delta_{t}-2}}-\frac{1}{R^{2\Delta_{t}-2}}\right\}.
\end{split}
\end{equation}
As $R\rightarrow \infty$ while keeping $c_t$ and cut-off $\epsilon$ invariant, the finite size invariant is $\langle M^\dagger M\rangle/N$. In Fig.~\ref{fig:order-parameter}, we show the finite size scaling of the rescaled quantity $\langle M^\dagger M\rangle/N$, and identify a crossing point around $v_0\in(-0.5, -0.3)$.

\section{S5: Magnetoroton analysis and projection}
\label{app:magnetorotor}

\subsection{Magnetoroton contribution to the low-energy spectrum}

The FQSH phase supports magnetoroton excitations~\cite{GMP1986} in both
$\mathcal M_T=+1$ and $-1$ sectors. At the decoupled point
$v_0=v_1=0$, exciting either of the two Laughlin layers gives
states of equal energy. Their combinations with definite
$\mathcal M_T$ form a degenerate pair of $\ell=2$ rotor states(For simplicity, we restrict our discussion to $\ell=2$ magnetorotons, since the low-energy states in this sector are most strongly affected by these excitations).
Below we display the overlaps only for $\mathcal M_T=+1$;
both sectors are included in the projection.
These charge-neutral collective excitations can occur in the
same energy range as the low-lying CFT states on a finite
sphere~\cite{Voinea_2025}. Their contribution is therefore
relevant to the larger finite-size corrections of the $\ell=2$
levels in the spectra presented in the main text. 

At the decoupled point, a density fluctuation acting on the FQSH
ground state gives a reference for the long-wavelength
magnetoroton excitation. In the $\ell=2$ sector we use
\begin{equation}
    |r_{\mathrm L}\rangle
    =\frac{\hat \rho_{2,0}|\Psi_{\mathrm{FQSH}}\rangle}
    {\|\hat \rho_{2,0}|\Psi_{\mathrm{FQSH}}\rangle\|},
    \qquad
    F_n=|\langle r_{\mathrm L}|\psi_n\rangle|,
    \label{eq:app-magnetorotor-overlap}
\end{equation}
where $|\Psi_{\mathrm{FQSH}}\rangle$ is the product of the two
opposite-chirality Laughlin states defined in the main text, and
$\hat\rho_{2,0}=\hat n^I_{2,0}$ is the $\ell=2,m=0$ component of
the total density $\hat n^I(\Omega)$. The state
$|\psi_n\rangle$ is the $n$th energy eigenstate in the
$(\ell,\mathcal M_T,\mathcal R_y)=(2,+,+)$ sector.

\begin{figure}[!ht]
    \centering
    \includegraphics[width=\textwidth]{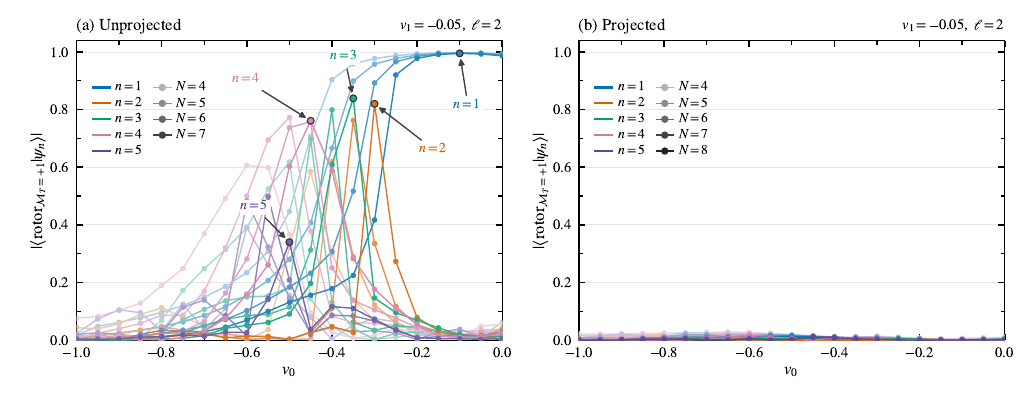}
    \caption{Magnetoroton overlaps along $v_1=-0.05$.
    (a) Overlap amplitudes $F_n$ for $N=4-7$, with the
    reference state $|r_{\mathrm L}\rangle$ defined in
    Eq.~\eqref{eq:app-magnetorotor-overlap}.
    (b) Residual overlap amplitudes
    $|\langle r_+|\psi_n^{(\lambda)}\rangle|$ after projection
    with $\lambda=8$, for $N=4-8$; $|r_+\rangle$ is the
    lowest decoupled $\ell=2$, $\mathcal M_T=+1$ eigenstate.
    Both panels show the five lowest levels in the
    $(\ell,\mathcal M_T,\mathcal R_y)=(2,+,+)$ sector on the
    same $y$-scale. Colors distinguish the energy rank $n$, and shades distinguish system sizes.
    Arrows in (a) mark the maximum overlap of each displayed
    $N=7$ level.}
    \label{fig:app-magnetorotor-overlap}
\end{figure}

Figure~\ref{fig:app-magnetorotor-overlap}(a) follows these overlaps
along $v_1=-0.05$. In the FQSH regime, the lowest $\ell=2$ state
has a large magnetoroton overlap. As the interlayer attraction
increases, the overlap transfers to higher excited states and
spreads over several levels near the transition. For example,
at $N=7$ the largest overlap moves from $n=1$ at $v_0=-0.20$
to $n=3$ at $v_0=-0.35$ and $n=4$ at $v_0=-0.45$.
Its amplitude at the transition is still $0.762$.
The low-energy $\ell=2$ spectrum therefore retains a substantial
magnetoroton contribution at the system sizes studied here.
This mixing can affect the comparison with the CFT operator
spectrum through the state-operator correspondence.

\subsection{Projection and the critical spectrum}

To isolate the effect of magnetoroton excitations, we project out the
lowest $\ell=2$ state of each symmetry sector at the decoupled
point. We denote the two normalized reference states by
$|r_\pm\rangle$, with $\mathcal M_T=\pm1$ and
$\mathcal R_y=+1$, and keep them fixed as $(v_0,v_1)$ varies.
The projection is implemented by adding a positive energy penalty,
\begin{equation}
    H_\lambda=H+\lambda P_r,
    \qquad
    P_r=|r_+\rangle\langle r_+|+|r_-\rangle\langle r_-|,
    \qquad \lambda=8,
    \label{eq:app-magnetorotor-penalty}
\end{equation}
in the $L_z=0$ Hilbert space. For $\lambda\to\infty$, the
low-energy states lie in the subspace orthogonal to both
references. We use $\lambda=8$, for which the residual overlaps
shown in Fig.~\ref{fig:app-magnetorotor-overlap}(b) are below
$0.029$ throughout the cut. Since $P_r$ has support only at
$\ell=2$, the $\ell=0,1$ spectra remain unchanged.
The exact diagonalization is performed for $N=4$--$8$ using the
SO3lver module of FuzzifiED~\cite{ZhouHe2026SO3lver}, retaining
the lowest twenty states in each $(\ell,\mathcal M_T)$ sector
for $\ell=0,1,2$.

\begin{figure}[!ht]
    \centering
    \includegraphics[width=\textwidth]{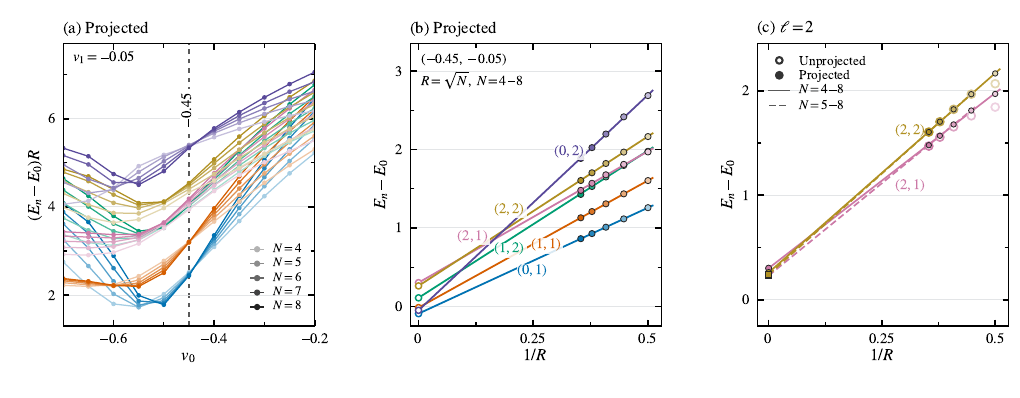}
    \caption{Finite-size scaling of the gaps after magnetoroton
    projection, with $R=\sqrt N$.
    (a) $(E_n-E_0)R$ along $v_1=-0.05$ for $N=4-8$.
    The dashed vertical line marks $v_0=-0.45$.
    (b) $E_n-E_0$ versus $1/R$ at $(v_0,v_1)=(-0.45,-0.05)$,
    with linear fits $E_n-E_0=a/R+b$ using $N=4-8$.
    $(\ell,n)$ label the levels in (a,b) follow
    Fig.~\ref{fig:scaled-gap-crossing}.
    (c) The lowest two gaps in the
    $(\ell,\mathcal M_T,\mathcal R_y)=(2,+,+)$ sector before
    (open symbols) and after (filled symbols) projection.
    Lines fit the projected data using $N=4-8$ (solid)
    and $N=5-8$ (dashed).}
    \label{fig:app-magnetorotor-gap}
\end{figure}

Fig.~\ref{fig:app-magnetorotor-gap} shows the gap scaling
after projection. The approximate collapse of the leading $\ell=0,1$ rescaled
gaps near $v_0=-0.45$ is unchanged. However, the $\ell=2$ gaps shift and their dependence
on $1/R$ becomes closer to linear
[Fig.~\ref{fig:app-magnetorotor-gap}(c)]. Fitting the projected
data to $a/R+b$ gives $b=0.302$ and $0.261$ for the two levels
using $N=4$--$8$, and $b=0.231$ and $0.242$ using $N=5$--$8$.
For comparison, the same fits before projection give
$\mathrm{gap}_\infty\equiv b=0.588$ and $0.492$ using $N=4$--$8$,
and $0.385$ and $0.240$ using $N=5$--$8$.
Projection reduces both intercepts, bringing the projected low-energy states closer to gapless.

\begin{figure}[!ht]
    \centering
    \includegraphics[width=\textwidth]{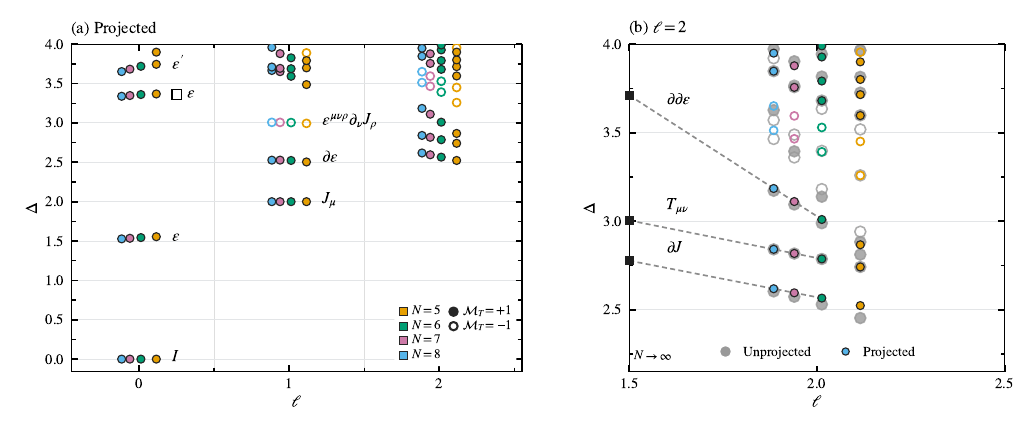}
    \caption{Operator spectra after magnetoroton
    projection at
    $(v_0,v_1)=(-0.45,-0.05)$ for $N=5-8$.
    (a) Operator spectra,
    fixing $\Delta_J=2$.
    Filled and open symbols denote $\mathcal M_T=+1$ and $-1$,
    respectively.
    (b) Enlarged view of the $\ell=2$ sector, with unprojected
    levels shown as gray circles and projected levels as colored symbols.
    Finite-size levels are shown for $\Delta<4$.
    The three lowest levels are labeled $\partial J$, $T_{\mu\nu}$,
    and $\partial\partial\epsilon$.
    Dashed lines in (b) fit $\Delta_\infty+b/N$ using $N=6,7,8$;
    squares mark the extrapolations.}
    \label{fig:app-magnetorotor-tower}
\end{figure}

Finally, Fig.~\ref{fig:app-magnetorotor-tower}(a) shows the effect of projection on the
operator spectra for $N=5$--$8$ at
$(v_0,v_1)=(-0.45,-0.05)$. The scalar and current levels at
$\ell=0,1$, including their descendant, are unaffected
by the projection. The change is confined to the $\ell=2$
sector, enlarged in Fig.~\ref{fig:app-magnetorotor-tower}(b).
Suppressing the magnetoroton contribution modifies the $\ell=2$ spectrum, while preserving the $\ell=0,1$
members of the scalar and current conformal families identified
in the main text. Fitting the three lowest projected $\ell=2$ levels to
$\Delta(N)=\Delta_\infty+b/N$ using $N=6,7,8$ gives the extrapolated
dimensions $\Delta_{\partial J}=2.777$, $\Delta_T=3.005$, and
$\Delta_{\partial\partial\epsilon}=3.711$.

\section{S6: Locating the critical line in the fermionic Coulomb bilayer}
\label{sec:critical}

\begin{figure}[bth]
  \centering
  \includegraphics[width=\linewidth]{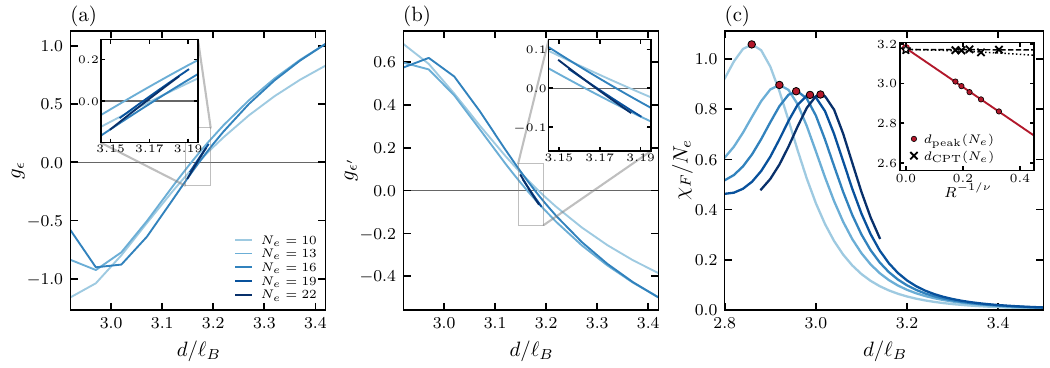}
  \caption{\label{fig:1323_111_cpt} Locating the critical point in the fermionic bilayer model at filling $(1/2, 2/3)$. Data is for fixed layer width $w = 2.5$ and system sizes $N_e = 10$--$22$ (light to dark). (a) Coupling $g_\varepsilon$ and (b) coupling $g_{\varepsilon'}$ from the first-order conformal perturbation theory of Eq.~\eqref{eq:cptfit}, plotted against $d$. The insets magnify the crossing region. The crossings $d_c(N_e)$ of $g_\varepsilon$ all lie in $3.157$--$3.173$, where $g_{\varepsilon'}$ is also close to zero. (c) Fidelity susceptibility per particle, $\chi_F/N_e$, with the peaks marked by red circles. Inset: peak positions $d_{\rm peak}(N_e)$ (circles) plotted against $R^{-1/\nu}$. We fit the peaks using $d_\infty + aR^{-1/\nu}$, giving $d_\infty^{\chi_F} = 3.178(4)$. Crosses mark the CPT roots $d_c(N_e)$, fitted to $d_\infty + (a \pm b)R^{-(1/\nu+\omega)}$ with the two signs corresponding to the parity of $N_e$ (dashed line is for $N_e$-even and plus sign; dotted line is for $N_e$-odd and minus sign), giving $d_\infty^{\rm CPT} = 3.172(5)$. }
\end{figure}

For the fermionic bilayer model introduced in End Matter, Fig.~\ref{fig:1323_111_cpt} locates the critical line $d_c(w)$ using two independent estimators, one based on the CFT spectrum and one on ground-state wavefunctions only. Along the entire critical line, the two estimators agree to within $0.01$ in $d_c$, as shown in Fig.~\ref{fig:1323_111_phasediagram} of End Matter.

{\em Conformal perturbation theory (CPT).---}Near criticality, the fuzzy sphere Hamiltonian can be viewed as the CFT perturbed by the relevant singlet $\varepsilon$ ($\Delta_\varepsilon = 1.51136$) and the leading irrelevant singlet $\varepsilon'$ ($\Delta_{\varepsilon'} = 3.794$), with couplings $g_\varepsilon$ and $g_{\varepsilon'}$~\cite{Lao2023}. To first
order, each scaling dimension measured, $\alpha = 2/(E_J - E_0)$, with the velocity fixed by $\Delta_J = 2$, is shifted in proportion to the
couplings,
\begin{equation}
  \alpha(E_O - E_0) - \Delta_O = g_\varepsilon f_{O\varepsilon O} + g_{\varepsilon'} f_{O\varepsilon' O}, \qquad O \in \{\varepsilon, \partial\varepsilon\},
  \label{eq:cptfit}
\end{equation}
where the coefficients are fixed by the O(2) OPE coefficients, $f_{\varepsilon V\varepsilon} = C_{\varepsilon V\varepsilon}$ and
$f_{\partial\varepsilon V\partial\varepsilon} = C_{\varepsilon V\varepsilon}\,[1 + \Delta_V(\Delta_V - 3)/(6\Delta_\varepsilon)]$ for the descendant, with
$C_{\varepsilon\varepsilon\varepsilon} = 0.8309$ and $C_{\varepsilon\varepsilon'\varepsilon} = 1.27$~\cite{Chester2020O2}. The two levels
$\varepsilon$ and $\partial\varepsilon$ give two equations, which we solve exactly for $(g_\varepsilon, g_{\varepsilon'})$ at every $(N_e, d)$. The
finite-size critical point $d_c(N_e)$ is the root where $g_\varepsilon$ changes sign from negative to positive. Its drift with $N_e$ is governed by the
leading irrelevant correction, $d_c(N_e) = d_\infty + (a \pm b)R^{-(1/\nu + \omega)}$ with $1/\nu + \omega = 2.283$, where the sign alternates with the
parity of $N_e$ to absorb the even/odd effect visible in the data. At $w = 2.5$ this gives $d_\infty^{\rm CPT} = 3.172$.
The second coupling carries useful information as well: $g_{\varepsilon'}$ at the critical point changes sign along the critical line at
$w^\ast \simeq 2.50$ [Fig.~\ref{fig:1323_111_cpt}(b)]. At this width the leading irrelevant perturbation is nearly tuned away, which is why we
adopt $w = 2.5$.

{\em Fidelity susceptibility.---}The overlap between neighbouring ground states, $F(d, \delta) = |\langle \psi_0(d) | \psi_0(d + \delta) \rangle|$, defines the fidelity
susceptibility~\cite{Zanardi2006,You2007,CamposVenuti2007} (for a review, see Ref.~\cite{Gu2010}) through $F = 1 - \tfrac12 \chi_F \delta^2 + O(\delta^3)$, with 
\begin{equation}
  \chi_F = \sum_{n \neq 0} \frac{|\langle n | \partial_d H | 0 \rangle|^2}{(E_n - E_0)^2}.
  \label{eq:chif}
\end{equation}
We evaluate it without computing excited states, as $\chi_F = \langle x | x \rangle$, where $|x\rangle$ is the first-order correction to the
ground state, $(H - E_0)|x\rangle = -(1 - |0\rangle\langle 0|)\,\partial_d H |0\rangle$. Near a continuous transition, $\chi_F$ is expected to develop a peak. At fixed $w$, we scan $d$, and the peak position approaches the critical point as $d_{\rm peak}(N_e) = d_\infty + a R^{-1/\nu}$~\cite{CamposVenuti2007,Gu2008, Albuquerque2010}, with
$\nu = 0.6717$~\cite{Chester2020O2}. Unlike CPT, this estimator uses no CFT data beyond $\nu$. At $w = 2.5$ the peaks extrapolate to
$d_\infty^{\chi_F} = 3.178$, which is close to the CPT value; we use $d = 3.18$. At the peak, the single state $\varepsilon$ carries
$96$--$97\%$ of the sum in Eq.~\eqref{eq:chif}. The transition is therefore driven by a single relevant singlet, as expected for O(2).

\section{Optimized Coulomb spectrum}
\label{app:optimised_111_1323}

\begin{figure}[h]
  \centering
  \includegraphics[width=\linewidth]{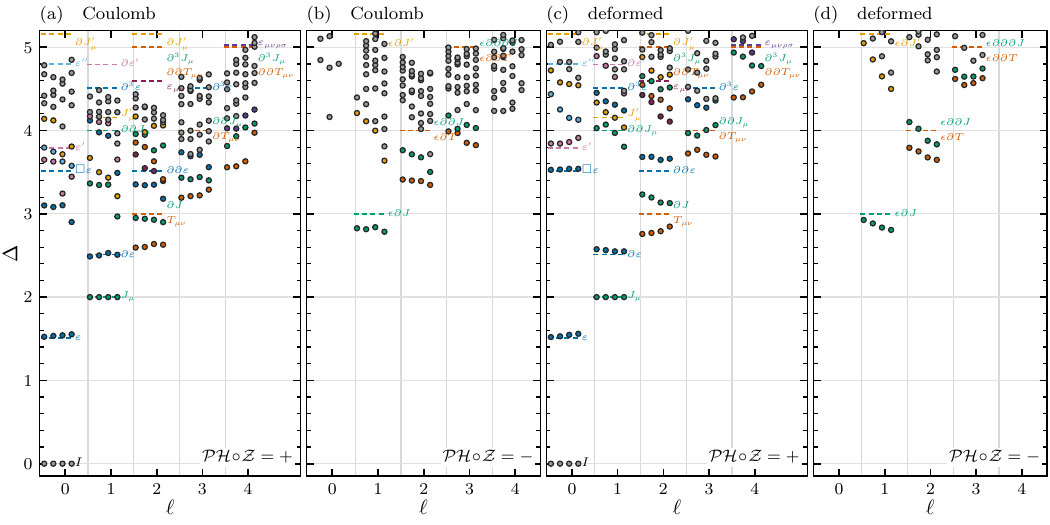}
  \caption{\label{fig:1323_111_optimised_spectrum} Neutral ($q = 0$) spectrum with the magnetorotons projected out (see text) for $N_e = 13$--$22$
(largest $N_e$ on the left in each column). (a)-(b) Bare Coulomb interaction at $w = 2.5$, $d = 3.18$, for $\mathcal{PH}\circ\mathcal{Z} = +$ and $-$. (c)-(d) The deformed interaction of Eq.~\eqref{eq:deformation} at
$\delta V_4^{\rm inter} = x_c$. Within each $(\ell,\mathcal{PH}\circ\mathcal{Z})$ sector, the $n$-th lowest level is compared with the $n$-th lowest O(2) scaling dimension
in that class (dashed lines), and is colored by the primary family ($\varepsilon$, $J_\mu$, $T_{\mu\nu}$, \dots) that this target belongs to;
grey states have no target in this window. Curl descendants are abbreviated by dropping indices, e.g.\ $\epsilon\partial J \equiv \epsilon^{\mu\nu\rho}\partial_\nu J_\rho$ and
$\epsilon\partial\partial T \equiv \epsilon^{\mu\nu\rho}\partial_\nu\partial_\alpha T_{\rho\lambda}$.}
\end{figure}

Figure~\ref{fig:1323_111_optimised_spectrum}(a)-(b) show the spectra of the bare Coulomb interaction with the previously identified optimal values of $w$ and $d$ in the two symmetry sectors of $\mathcal{PH}\circ\mathcal{Z}$. In both cases we explicitly project out the magnetoroton excitations that do not belong to the CFT spectrum, as explained previously. Specifically,  at each $\ell$, the lowest level of the decoupled
Hamiltonian, $H_{\rm dec} = H^{\rm intra}_1 + H^{\rm intra}_2$ (i.e., keep the same intralayer interaction and set $V^{\rm inter} = 0$), is a degenerate pair (a roton in either layer), since the $2/3$ layer is the particle--hole conjugate of
the $1/3$ layer. Its two $\mathcal{PH}\circ\mathcal{Z}$ combinations $r_\pm$ are kept fixed and penalized, $H \to H + \lambda(|r_+\rangle\langle r_+| + |r_-\rangle\langle r_-|)$ with a large coupling $\lambda = 8\,e^2/\epsilon\ell_B$, so each class loses one state per $\ell$.

The bare Coulomb spectrum [Fig.~\ref{fig:1323_111_optimised_spectrum}(a, b)] reproduces the lowest O(2) operators, but $T_{\mu\nu}$ is too low and extra non-roton levels appear above $\Delta \approx 2.5$. Such deviations come from irrelevant operators with large couplings, so one should be able to tune them away by a small change to the short-range part of the interaction. To test this, we slightly deform the two shortest-range intralayer pseudopotentials of both layers,
\begin{equation}
  V_m^{\uparrow\uparrow} \to V_m^{\uparrow\uparrow} + \delta V_m^{\uparrow\uparrow}, \qquad (\delta V_1^{\uparrow\uparrow}, \delta V_3^{\uparrow\uparrow}) = (-0.0101, -0.0063),
  \label{eq:deformation}
\end{equation}
and similarly for $V_m^{\downarrow\downarrow}$. This is a small change amounting to about $3\%$ of $V_1^{\uparrow\uparrow}$ and $V_3^{\uparrow\uparrow}$. The shifts are fixed by a fit to the vacuum spectrum:
the six lowest levels ($\varepsilon$, $J_\mu$, $\partial\varepsilon$, $\epsilon^{\mu\nu\rho}\partial_\nu J_\rho$, $\partial J$ and $T_{\mu\nu}$), assigned
by rank within each $(\ell, \mathcal{PH}\circ\mathcal{Z})$ sector, are fitted jointly at $N_e = 10, 13, 16$ with one velocity $\alpha_{N_e} \propto R_{N_e}$ shared
by all sizes. The common velocity forces the gaps to close as $1/R$, which excludes gapped points that mimic the CFT ratios size by size.
The levels $\Box\varepsilon$, $\varepsilon'$ and all higher states are not fitted and serve as checks.

The deformation in Eq.~(\ref{eq:deformation}) moves the system slightly off the critical line. We restore criticality with a single interlayer shift $\delta V_4^{\uparrow\downarrow} = x$ (below $0.5\%$ of $V_4^{\uparrow\downarrow}$), locating $x_c$ from the $\chi_F$ peak in $x$ as in Appendix~\ref{sec:critical}. Here
the peak drift is not dominated by the leading $R^{-1/\nu}$ term, and the fit including the subleading correction in $R^{-1/\nu+\omega}$ gives
$x_c = -6.8(1.5) \times 10^{-4}$; the leading form alone gives $-3.2 \times 10^{-3}$ where the spectrum does not look critical, so $x_c$ is less certain than $d_c$ at the Coulomb point. At this value $x_c$,
the low-lying neutral spectrum is much closer to that of the O(2) CFT, as seen in Fig.~\ref{fig:1323_111_optimised_spectrum}(c)-(d). For example, at system size $N_e = 22$,
$\Delta_\varepsilon = 1.520$ (O(2) expectation: 1.511), $\epsilon^{\mu\nu\rho}\partial_\nu J_\rho$ lies at $2.93$ (expected 3) and $T_{\mu\nu}$ at $2.75$ (expected 3). The unfitted
$\Box\varepsilon$ at $3.53$ (3.511) and $\varepsilon'$ at $3.84$ (3.794) now sit at their targets instead of being displaced by extra levels. The rms
deviation of the eight lowest levels falls from $0.24$--$0.31$ to $0.12$--$0.16$ ($N_e = 13$--$22$), and the layer-odd magnetoroton rises from
$\Delta \approx 2.5$ to $\approx 3.1$--$3.3$.

The deformation thus optimizes the vacuum sector but not the others. Its criticality is supported by the $\chi_F$ peak along
$\delta V_4^{\rm inter}$ and by $\Delta_\varepsilon$ approaching the O(2) value ($1.547$, $1.529$, $1.520$ at $N_e = 16, 19, 22$), but the
charged and fractionalized sectors move further from the scaling regime: the $Q = -1$ descendant spacing $\Delta_{\partial\chi} -
\Delta_\chi$, which should equal 1, grows from $0.82$ to $1.34$ over $N_e = 10$--$22$, and the even- and odd-$\ell$ terms of
$G(\pi) = \sum_\ell (-1)^\ell \frac{2\ell+1}{4\pi} c_\ell$ ($c_\ell \ge 0$ the weight of $\Phi_\ell$ on the $Q = -1$ states at angular
momentum $\ell$) nearly cancel, so that $\hat G_-(\pi)$ changes sign with $N_e$ and $\Delta_\chi$ cannot be extracted. Restoring criticality
instead with $d$ or $\delta V_0^{\rm inter}$, using either form of the $\chi_F$ extrapolation, gives a positive $\hat G_-(\pi)$ at all $N_e$
but a markedly worse vacuum spectrum (rms deviation $0.22$--$0.34$, $\Delta_\varepsilon$ up to $1.65$), so $\delta V_4^{\rm inter}$ is the
choice that best optimizes the vacuum. Tuning two intralayer pseudopotentials is therefore not enough to bring all sectors to the CFT
simultaneously; a broader deformation involving further intra- and interlayer pseudopotentials may achieve this, but we have not explored
it. None of the XY$^\ast$ evidence relies on the deformed point: the charged sectors and the exciton dimension are taken at the plain
Coulomb point, whose criticality is established independently.

\section{S7: Pair tunneling and Ising\texorpdfstring{$^*$}{*} criticality}
\label{app:ising-star}

\subsection{Pair-tunneling Hamiltonian}
In the main text, we focus on the fractionalized XY$^*$ transition. Within a similar setup, 
the bilayer model admits an
Ising$^*$ transition under pair tunneling. For two layers with opposite magnetic flux, we use the rotationally invariant Hamiltonian
\begin{align}
    H &= H_{\mathrm{intra}}+H_{\mathrm{pair}}
       +\mu(\hat{\mathcal N}-2N),
       \label{eq:app-ising-hamiltonian}\\
    H_{\mathrm{pair}} &= t\sum_{m=-s}^{s}(-1)^{s+m}
    \left(b^\dagger_{m,\uparrow}b^\dagger_{-m,\downarrow}
    +b_{-m,\downarrow}b_{m,\uparrow}\right).
    \label{eq:app-ising-pair}
\end{align}
It is worth noting that the pair-tunneling term breaks $U(1)_c$ symmetry explicitly. 
We retain the intralayer Haldane pseudopotentials $v_0=1$ and $v_2$, set the remaining pseudopotentials and interlayer
interaction to zero, and vary $(v_2,t)$ at fixed $\mu=-2.1$.
Energies are measured in units of $v_0$.
The reference particle number per layer $N$ fixes the flux,
$2s=2N-2$. We diagonalize in the balanced sector
$N_\uparrow-N_\downarrow=0$, with total particle number
$\mathcal N$ taking the values
$2N-\Delta N,2N-\Delta N+2,\ldots,2N+\Delta N$.
The pair term preserves total particle-number parity and the
relative $U(1)$ symmetry. We label the levels by angular momentum
$\ell$ and the time-reversal quantum number $\mathcal M_T$
in the real basis.

Since the interlayer pair operator satisfies $B\sim\phi^2$, the pair
tunneling adds $\lambda(\phi^2+\phi^{*2})$ to the effective theory.
Writing $\phi=(\varphi_1+i\varphi_2)/\sqrt{2}$ gives
\begin{equation}
    V_2=\frac{r+2\lambda}{2}\varphi_1^2
       +\frac{r-2\lambda}{2}\varphi_2^2.
    \label{eq:app-ising-anisotropy}
\end{equation}
Pair tunneling splits the masses of the two real components.
Integrating out the massive component near the transition leaves
a real scalar $\varphi$ with a quartic interaction. This field
describes the bosonic bound state of a semion and an antisemion
and carries $\mathbb Z_2$ gauge charge. Its condensation confines
the semion and antisemion, yielding a topologically trivial phase. The local critical
operators are even under $\varphi\to-\varphi$, giving the
Ising$^*$ operator content~\cite{Schuler2016Torus}.

\subsection{Ground-state evolution}

\begin{figure}[!ht]
    \centering
    \includegraphics[width=\textwidth]{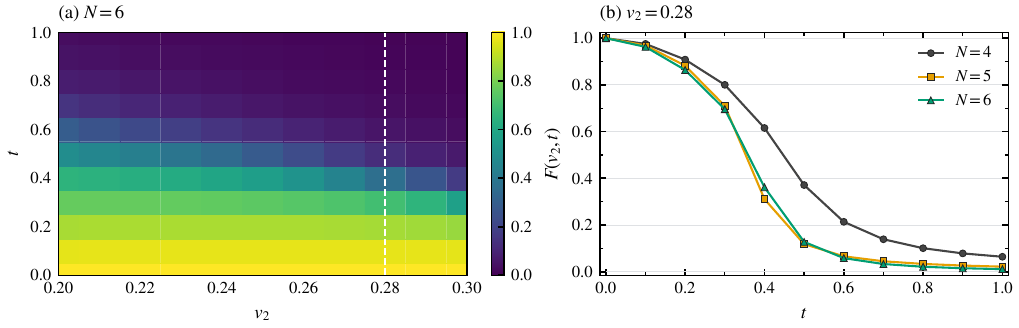}
    \caption{Ground-state fidelity $F(v_2,t)$ at $\Delta N=2$.
    (a) Parameter scan for $N=6$, with the $v_2=0.28$ cut
    marked by the dashed line.
    (b) The same cut for $N=4,5,6$.}
    \label{fig:app-ising-fidelity}
\end{figure}

We track the ground-state evolution at $\Delta N=2$ through
the fidelity with the $t=0$ ground state at the same $v_2$,
\begin{equation}
    F(v_2,t)=
    \left|\langle\Psi_0(v_2,0)|\Psi_0(v_2,t)\rangle\right|^2.
    \label{eq:app-ising-fidelity}
\end{equation}
The ground state has $\mathcal M_T=+1$ throughout the scan.
The fidelity decreases as $t$ increases
[Fig.~\ref{fig:app-ising-fidelity}]. At $(v_2,t)=(0.28,0.5)$,
it is $0.371$, $0.118$, and $0.128$ for $N=4,5,6$, respectively.
By $t=1$, all three values are below $0.07$.

\FloatBarrier

\subsection{Low-energy spectra}

We compare the low-energy spectrum with the leading even scalar
$\epsilon$ of the three-dimensional Ising CFT and its descendants.
For this matching, we assign the first two excited
$(\ell,\mathcal M_T)=(0,+)$ levels to $\epsilon$ and
$\Box\epsilon$, and the lowest $(1,+)$ level to
$\partial\epsilon$. Their Ising dimensions are $\Delta_\epsilon$,
$\Delta_\epsilon+2$, and $\Delta_\epsilon+1$, respectively, with
$\Delta_\epsilon\simeq1.413$~\cite{Chang2025IsingStressTensor}.
With $E_0$ the ground-state energy, we normalize the gaps as
\begin{equation}
    \Delta_a=1.413\frac{E_a-E_0}{E_\epsilon-E_0}
    \label{eq:app-ising-normalization}
\end{equation}
and minimize the sum of the descendant deviations,
\begin{equation}
    f=|\Delta_{\partial\epsilon}-2.413|
      +|\Delta_{\Box\epsilon}-3.413|.
    \label{eq:app-ising-cost}
\end{equation}

\begin{figure}[!ht]
    \centering
    \includegraphics[width=\textwidth]{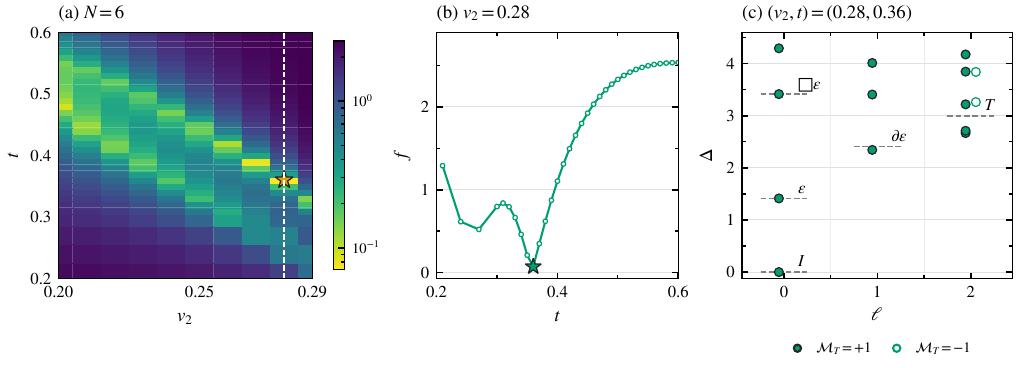}
    \caption{Spectral matching for $N=6$ at $\Delta N=4$.
    (a) Cost $f$ for $t\geq0.21$, on a logarithmic color scale.
    (b) Cut along $t$ at $v_2=0.28$.
    Stars mark the minimum at $(v_2,t)=(0.28,0.36)$.
    (c) Spectrum at the minimum, showing $0\leq\ell\leq2$
    and $\Delta<4.5$ using Eq.~\eqref{eq:app-ising-normalization}.
    Filled and open circles denote $\mathcal M_T=+1$ and $-1$;
    horizontal offsets separate the two sectors.
    Dashed segments give the Ising reference dimensions of
    $I$, $\epsilon$, $\partial\epsilon$, $\Box\epsilon$,
    and the stress tensor $T$ with $\Delta_T=3$.}
    \label{fig:app-ising-cost}
\end{figure}

For $N=6$ and $\Delta N=4$, we scan
$0.20\leq v_2\leq0.29$ and $0.03\leq t\leq0.60$.
The smallest cost occurs at $(v_2,t)=(0.28,0.36)$, with
$f=0.07108$, $\Delta_{\partial\epsilon}=2.34398$, and
$\Delta_{\Box\epsilon}=3.41506$.
The spectrum at this minimum is shown for $0\leq\ell\leq2$
in Fig.~\ref{fig:app-ising-cost}(c).
This optimized operator spectrum shows a clear feature of Ising$^*$ CFT, where $\epsilon$-conformal family can be clearly seen.  

\end{document}